\documentclass[onecolumn,10pt,showpacs,amsmath,amssymb]{revtex4}
\usepackage[cp866]{inputenc}
\usepackage[english]{babel}
\usepackage{graphicx}
\usepackage{bm}

\newcommand{\sgn}{{\rm sgn}}

\begin{document}
\title{Shear viscosity in neutron star cores}

\author{P.~S.\ Shternin 
and D.~G.\ Yakovlev }

\affiliation{Ioffe Physical Technical Institute,
Politekhnicheskaya 26, 194021 Saint-Petersburg, Russia}

\date{\today}

\begin{abstract}
We calculate the  shear viscosity $\eta \approx
\eta_\mathrm{e\mu}+\eta_\mathrm{n}$ in a neutron star core
composed of nucleons, electrons and muons ($\eta_\mathrm{e\mu}$
being the electron-muon viscosity, mediated by collisions of electrons
and muons with
charged particles, and $\eta_\mathrm{n}$ the neutron viscosity,
mediated by neutron-neutron and neutron-proton collisions).
Deriving $\eta_\mathrm{e\mu}$, we take into account the Landau
damping in collisions of electrons and muons with charged
particles via the exchange of transverse plasmons. It lowers
$\eta_\mathrm{e\mu}$ and leads to the non-standard temperature behavior
$\eta_\mathrm{e\mu}\propto T^{-5/3}$. The viscosity $\eta_{\rm n}$
is calculated taking into account that in-medium effects modify
nucleon effective masses in dense matter.
Both viscosities, $\eta_\mathrm{e\mu}$
and $\eta_\mathrm{n}$, can be important, and both are calculated
including the effects of proton superfluidity. They are presented
in the form valid for any equation of state of nucleon dense
matter. We analyze the density and temperature dependence of
$\eta$ for different equations of state in neutron star cores, and
compare $\eta$ with the bulk viscosity in the core and with the
shear viscosity in the crust.
\end{abstract}


\maketitle

\section{Introduction}
\label{introduc}

Neutron stars are very compact. Their typical masses are $\sim
1.4\, M_\odot$ (where $M_\odot$ is the mass of the Sun), while
their radii are as small as $\sim 10$~km. As a result, a neutron
star core contains matter, whose density $\rho$
reaches several $\rho_0$ ($\rho_0 \approx 2.8 \times 10^{14}$ g~cm$^{-3}$
being the density of the standard saturated nuclear matter). The
core is composed of uniform neutron-rich nuclear
matter and extends from $\rho \approx 0.5\,\rho_0$ to the stellar
center (where $\rho$ can be as high as $10 \rho_0$).
It attracts special attention because of its
poorly known composition and equation of state (EOS);
e.g., Ref.\ \cite{hpy07}. From
outside,  the core is surrounded by a thin ($\sim 1$
km thick) and light (a few per cent by mass) crust composed of
atomic nuclei, strongly degenerate electrons and (after the
neutron drip at $\rho \gtrsim 4 \times 10^{11}$ g~cm$^{-3}$) free
neutrons.

In this paper, we study the shear viscosity of neutron
star cores. It is an important transport property
which affects the relaxation of hydrodynamic motions,
particularly, a possible differential rotation
within the star and
stellar oscillations \cite{cl87}.  The shear viscosity can be
important for damping gravitational wave driven instabilities
(for instance, r-modes;
e.g., \cite{anderssonetal00} and references therein).
Its knowledge is required
to analyze the efficiency of
such instabilities for generating gravitational waves.

For simplicity, we consider the cores composed of strongly
degenerate neutrons (n), protons (p), electrons (e), and muons
($\mu$) -- $\mathrm{npe\mu}$-matter, neglecting a possible
appearance of hyperons and/or exotic forms of matter (pion or kaon
condensates or quarks or their mixtures) as predicted by some EOSs
at $\rho \gtrsim 2\,\rho_0$; see, e.g., Ref.\ \cite{hpy07}.
The electrons and muons
constitute almost ideal gases.
The muons are absent in the
outermost part of the core. They appear at densities exceeding a
threshold value $\rho_\mu \sim \rho_0$ \cite{st83} at which the
electron chemical potential reaches the muon
rest-mass energy ($\mu_{\rm e}=m_\mu c^2 \approx 207\,m_{\rm
e}c^2$). The electrons are ultra-relativistic,
while the muons are non-relativistic just
after the threshold but become relativistic at higher $\rho$. In
contrast to electrons and muons, nucleons constitute a strongly
interacting Fermi liquid where protons are essentially
non-relativistic, while neutrons become mildly relativistic at
$\rho \gtrsim 2\,\rho_0$. The neutrons and protons can be in
superfluid state (e.g., Ref.\ \cite{ls01}).

The main contribution to the shear viscosity $\eta$
in a neutron star core comes from electrons and
muons (lightest and most mobile particles) and neutrons (most
abundant particles),
\begin{equation}
  \eta=\eta_\mathrm{e\mu}+\eta_\mathrm{n}.
\label{eta_emun}
\end{equation}
The viscosity
$\eta_\mathrm{e\mu}$ of electrons and muons is mainly limited by
collisions of electrons and muons between themselves
and with other charged particles (protons, in our case)
via electromagnetic forces. In contrast, the
neutron contribution $\eta_\mathrm{n}$ is limited by
neutron-neutron and neutron-proton collisions mediated by
strong interactions. As a result, $\eta_\mathrm{e\mu}$
and $\eta_\mathrm{n}$ are nearly independent
(belong to different -- electromagnetic and
nuclear -- sectors) and can be
calculated separately \cite{fi79}.

In applications, one often employs the viscosity
$\eta_\mathrm{e\mu}$ calculated by Flowers and Itoh \cite{fi79}
for non-superfluid matter. Recently, Andersson et al.\
\cite{acg05} have estimated $\eta_\mathrm{e\mu}$
for superfluid matter. However, these studies neglect an
enhancement of collisions of relativistic charged
particles due to the exchange of transverse plasmons. The
significance of this effect was demonstrated by Heiselberg and
Pethick \cite{hp93} in their study of transport properties of
ultra-relativistic quark matter. Recently we (Shternin and Yakovlev
\cite{sy07} -- hereafter  SY07) have reconsidered the
electron-muon thermal conductivity $\kappa_\mathrm{e\mu}$ taking
into account the exchange of transverse plasmons. This effect can
reduce $\kappa_\mathrm{e\mu}$ by several orders of magnitude.

Here we reanalyze $\eta_\mathrm{e\mu}$ in the same manner.
We closely follow SY07 and omit
technical details.
In addition, we reconsider $\eta_\mathrm{n}$, which is a more
difficult task involving
nucleon-nucleon collisions.
The viscosity $\eta_\mathrm{n}$  was calculated by Flowers and Itoh
\cite{fi79} for one EOS of non-superfluid
matter assuming in-vacuum nucleon-nucleon scattering.
These results were fitted
by Cutler and Lindblom \cite{cl87} by a simple analytical
expression which is widely used;
according to Ref.\ \cite{fi79},
 $\eta_\mathrm{n}>\eta_\mathrm{e\mu}$. Recently Benhar and Valli
\cite{bv07} have calculated $\eta_\mathrm{n}$ for pure neutron
matter in a self-consistent manner using the same nucleon interaction
potential to derive $\eta_\mathrm{n}$ and construct the EOS (also
for one EOS).
We calculate
$\eta_\mathrm{n}$ in a more general way
than Flowers and Itoh
\cite{fi79}. Our approach is similar to that used
by Baiko, Haensel and Yakovlev \cite{bhy01} (hereafter BHY01) for
evaluating the thermal conductivity of neutrons.
In addition, we employ recent developments \cite{zhangetal07} in
calculations of nucleon-nucleon scattering
cross sections in nuclear matter. As in BHY01,
we take into account superfluidity of protons.
Again, we closely follow the derivation of BHY01 and omit
the details.

After calculating $\eta_\mathrm{e\mu}$ and $\eta_\mathrm{n}$, we
analyze the shear viscosity in
neutron star cores with different EOSs.

\section{Shear viscosity in non-superfluid matter}
\label{formalism}

The shear viscosity is calculated from a system of
coupled Boltzmann kinetic equations
\begin{equation}
\label{E:KinEqGeneral}
  \bm{v}_c\,\frac{\partial F_c}{\partial \bm{r}}=\sum_{i} I_{ci},
\end{equation}
where $F_c$ is the distribution function of
momentum-transfer carriers $c$ (with $c=e$,
$\mu$, or $n$, in our case);
$i=$n, p, e, $\mu$ runs over all particle species;
$\bm{v}_c$ is the velocity of particles $c$, and $I_{ci}$ is a
collision integral, that describes a scattering of
particles $c$ and $i$:
\begin{eqnarray}
    I_{ci}&=&
     \frac{1}{(2\pi\hbar)^9(1+\delta_{ci})}
     \sum_{\sigma_{1'}\sigma_2\sigma_{2'}}\,\int
     {\rm d}\bm{p}_2\, {\rm d}\bm{p}_{1'}\, {\rm d}
    \bm{p}_{2'}\; w_{ci}(12|1'2')
\nonumber \\
    & & \times \left[F_{1'}F_{2'}(1-F_1)(1-F_2)-F_1F_2(1-F_{1'})(1-F_{2'})\right].
\label{collint}
\end{eqnarray}
Here, 1 and 2 denote particle states before a collision; 1$'$ and
2$'$ are particle states after the collision; $\bm{p}$ is the
particle momentum, $\sigma$ is the spin state, and $w_{ci}$ is the
differential transition probability. The Kronecker delta $\delta_{ci}$ is
included to avoid double counting of collisions between
identical particles ($c=i$).

Distributions $F_c$ slightly deviate from the equilibrium
Fermi-Dirac distributions $f_c$ owing to the presence of a
small hydrodynamical velocity field $\bm{V}$,
\begin{equation}
    F_c=f_c -\Phi_c\,
    {\partial f_c  \over \partial \varepsilon_c},
    \quad f_c= \left\{ \exp \left( {\varepsilon_c - \mu_c \over
    k_B T} \right) +1 \right\}^{-1},
\label{distributions}
\end{equation}
where $\varepsilon_c$ is the particle energy, $\mu_c$ is its
chemical potential, $T$ is the temperature, $k_B$ is the Boltzmann
constant, and $\Phi_c$  measures a deviation from equilibrium. The
electron-muon and neutron transports are decoupled because we
neglect electromagnetic interaction between the leptons and
neutrons. For calculating $\eta_\mathrm{e\mu}$, the electrons and
muons are treated as the only momentum carriers which undergo
collisions between themselves and with protons. For calculating
$\eta_\mathrm{n}$, the only momentum carriers are assumed to be
neutrons, while the contribution of protons is neglected due to
their small fraction. Therefore, the protons are thought to be
passive scatterers which obey the equilibrium Fermi-Dirac
distribution.
Nonequilibrium parts
of the electron, muon, and neutron distributions are found using
the standard variational approach with the simplest trial
function,
\begin{equation}
     \Phi_c= -\tau_c \,
     \left(v_{c\alpha}p_{c\beta}
     -\frac{1}{3}\,v_cp_c\delta_{\alpha\beta}\right)V_{\alpha\beta},
\label{Phi}
\end{equation}
where $\tau_c$ is an effective relaxation time of particles
$c$, $\bm{v}_c$ is their velocity, and
\begin{equation}
   V_{\alpha\beta}=\frac{1}{2}\left(\frac{\partial V_\alpha}{\partial
    x_\beta}+\frac{\partial V_\beta}{\partial x_\alpha}\right),
\end{equation}
with $\sum_\alpha V_{\alpha\alpha}={\rm div}\bm{V}=0$.

The resulting shear viscosity
is expressed through the effective
relaxation times in a standard way,
\begin{equation}
\label{eta}
     \eta=\eta_\mathrm{e\mu}+\eta_\mathrm{n}
     =\eta_\mathrm{e}+\eta_{\mu}+\eta_\mathrm{n},\ \ \ \ \
     \eta_\mathrm{e}=
     \frac{n_\mathrm{e} p_{F \rm e}^2 \tau_\mathrm{e}}{5m_\mathrm{e}^*},\
     \eta_\mu=\frac{n_\mu p_{F\mu}^2 \tau_\mu}{5m_\mu^*},
     \ \eta_\mathrm{n}=
     \frac{n_\mathrm{n} p_{F\rm n}^2\tau_\mathrm{n}}{5m_\mathrm{n}^*},
\end{equation}
where $\eta_\mathrm{e}$, $\eta_\mu$, and $\eta_\mathrm{n}$ are,
respectively, the partial electron, muon, and neutron shear
viscosities; $n_c$ is the number density of particles $c$;
$p_{Fc}$ is their Fermi momentum; and $m_c^*$ is an effective
mass on their Fermi surface. The electron and muon
effective masses differ from their rest masses due to relativistic
effects, $m_\mathrm{e}^*=\mu_\mathrm{e}/c^2$ and
$m_\mu^*=\mu_\mu/c^2$. The neutron and proton effective masses
differ from 
their bare masses
mainly due to many-body effects in
dense matter (being determined by neutron and proton densities of state near
appropriate Fermi surfaces).

Linearizing the kinetic equations, multiplying them by
$\left(v_{1\alpha}p_{1\beta}-\frac{1}{3}v_1p_1\delta_{\alpha\beta}\right)$,
summing over $\sigma_1$ and integrating over
$(2\pi\hbar)^{-3}{\rm d}\bm{p}_1$, we obtain
a system of equations for the relaxation times,
\begin{equation}
     1=\sum_i \left( \nu_{ci}\tau_c+ \nu_{ci}'\tau_i\right),\ \ \ \
     c=\rm{e},\ \mu,\ \rm{n},
\label{eq:coll_syst}
\end{equation}
where we introduce the effective collision frequencies,
\begin{eqnarray}
      \nu_{ci}&=&\frac{3\pi^2\hbar^3}{2 p_{Fc}^5 k_BT m_c^*}\int
      \frac{{\rm d} \bm{p}_1 \,{\rm d} \bm{p}_{1'}\,
      {\rm d}\bm{p}_2\,{\rm d} \bm{p}_{2'}}{(2\pi\hbar)^{12}}\;
      W_{ci}(12|1'2')\, f_1f_2(1-f_{1'})(1-f_{2'})
\nonumber\\
      &\times& \left[ \frac{2}{3} \,p_1^4+
      \frac{1}{3}\, p_{1}^2p_{1'}^2-
      (\bm{p}_1\cdot \bm{p}_{1'})^2\right],
\label{eq:nuci_int}\\
       \nu_{ci}'&=&\frac{3\pi^2\hbar^3}{2 p_{Fc}^5 k_BT m_i^*}\int
       \frac{{\rm d} \bm{p}_1 \,{\rm d} \bm{p}_{1'} \,{\rm d}
       \bm{p}_2\, {\rm d} \bm{p}_{2'}}{(2\pi\hbar)^{12}}\,
       W_{ci}(12|1'2')\; f_1f_2(1-f_{1'})(1-f_{2'})
\nonumber\\
       &\times&
       \left[\frac{1}{3}\,p_1^2p_{2'}^2-\frac{1}{3}\,
       p_1^2p_2^2+(\bm{p}_1 \cdot \bm{p}_2)^2-
       (\bm{p}_1\cdot\bm{p}_{2'})^2\right],
\label{eq:nuci1_int}
\end{eqnarray}
with
$W_{ci}(12|1'2')=(1+\delta_{ci})^{-1}
\sum\limits_\mathrm{spins}w_{ci}(12|1'2')$
(the sum is over spin states of all particles 1,2,1$'$,2$'$).

The formal solution of (\ref{eq:coll_syst})
for the $\mathrm{npe\mu}$-matter is
%
\begin{equation}
   \tau_{\rm e} =
   {\nu_\mu -\nu'_{\rm e\mu} \over \nu_{\rm e} \nu_\mu
   - \nu'_\mathrm{e\mu}\nu'_\mathrm{\mu e}},
   \quad
   \tau_\mu =
   {\nu_{\rm e} -\nu'_\mathrm{\mu e} \over
   \nu_{\rm e} \nu_\mu - \nu'_\mathrm{e\mu}
   \nu'_\mathrm{\mu e}},\quad
   \tau_\mathrm{n}={1 \over \nu_\mathrm{n}},
\label{taus}
\end{equation}
where
\begin{eqnarray}
     \nu_{\rm e} &=& \sum_i \nu_{{\rm e}i}+\nu_\mathrm{ee}'
     =\nu_\mathrm{ee}+\nu_\mathrm{ee}'
     +\nu_\mathrm{e \mu}+\nu_\mathrm{ep},
\nonumber \\
     \nu_\mu&=& \sum_i \nu_{\mu i}+\nu_{\mu\mu}'
     =\nu_{\mu \mu}+\nu_{\mu \mu}'+
     \nu_\mathrm{\mu e}+ \nu_\mathrm{\mu p},
\nonumber\\
      \nu_\mathrm{n}&=&\nu_\mathrm{nn}+\nu_\mathrm{nn}'+\nu_\mathrm{np}.
\label{nus}
\end{eqnarray}
In the absence of muons, the expression for $\eta_\mathrm{e\mu}$
simplifies,
\begin{equation}
   \eta_\mathrm{e\mu}=\eta_\mathrm{e},
   \quad  \tau_\mathrm{e}^{-1}=
   \nu_{\rm e}= \nu_\mathrm{ee}+\nu_\mathrm{ee}'+\nu_\mathrm{ep}.
\label{nomuons}
\end{equation}
%

Once collision frequencies are found, the viscosity is
obtained from Eq.\ (\ref{eta}). In order to determine the
collision frequencies from Eqs.\ (\ref{eq:nuci_int}) and
(\ref{eq:nuci1_int}) one needs to know the
transition probability $W_{ci}(12|1'2')$.
The collisions of charged particles
should be considered with
a proper treatment of plasma
screening of electromagnetic
interaction. We discuss the plasma screening
and the calculation of $\eta_\mathrm{e\mu}$ in
Secs.\ \ref{viscosity}--\ref{corr}. The neutron viscosity $\eta_\mathrm{n}$
is studied in Sec.~\ref{viscn_norm}.
In Sec.\ \ref{formalism} we consider nonsuperfluid
nucleons; the effects of proton superfluidity are
analyzed in Sec.\ \ref{superflu}.
Throughout the paper we use the simplest variational approach. A
comparison with an exact solution is discussed in Sec.\ \ref{S:exact}.



\subsection{Plasma screening}
\label{viscosity}

The plasma screening in neutron star
cores is discussed in SY07. Here, we outline the main points.

The differential collision probability can be written as
\begin{equation}
\label{eq:Wci}
     W_{ci}(12|1'2')=4\frac{(2\pi\hbar)^4}{\hbar^2}
     \,\delta(\bm{p}_1+\bm{p}_2-\bm{p}_{1'}-\bm{p}_{2'})
     \, \delta(\varepsilon_1+\varepsilon_2-\varepsilon_{1'}-\varepsilon_{2'})
     \,\frac{\langle|M_{ci}|^2\rangle}{1+\delta_{ci}},
\end{equation}
where $\langle|M_{ci}|^2\rangle$ is the squared matrix
element summed over final and averaged over initial spin states.
For collisions of identical particles,
we have $M_{cc}=M_{cc}^{(1)}-M_{cc}^{(2)}$, where
the first and second terms correspond to the scattering channels
$12\to 1'2'$ and $12\to 2'1'$,
respectively. Collisions of
different particles go through a single channel, $M_{ci}=M_{ci}^{(1)}$,
\begin{equation}
    M_{ci}^{(1)}
    =\frac{4\pi e^2}{c^2}
    \left(\frac{J_{1'1}^{(0)}J_{2'2}^{(0)}}{q^2+\Pi_l}-
     \frac{\bm{J}_{t1'1}\cdot\bm{J}_{t2'2}}{q^2-\omega^2/c^2+\Pi_t}\right),
\label{Mfi}
\end{equation}
where $\hbar \bm{q}=\bm{p}_{1'}-\bm{p}_1$ and $\hbar
\omega=\varepsilon_{1'}-\varepsilon_1$ are, respectively, momentum
and energy transfers in a collision event;
$J_{c'c}^{(\nu)}=(J_{c'c}^{(0)},
\bm{J}_{c'c})=(2m_c^*c)^{-1}(\bar{u}_{c'}\gamma^\nu u_c)$ is the
transition 4-current ($\nu=0$, 1, 2, 3), $\bm{J}_{tc'c}$ is the
component of $\bm{J}_{c'c}$ transverse to $\bm{q}$; $\gamma^\nu$
is a Dirac matrix; $u_c$ a normalized  bispinor (with
$\bar{u}_cu_c=2m_cc^2$), and $\bar{u}_c$ is a Dirac conjugate. The
first term in Eq.\ (\ref{Mfi}) corresponds to direct Coulomb
interaction via the longitudinal currents (with respect to
$\bm{q}$); the space-like longitudinal component of the current is
expressed through the time-like component $J_{c'c}^{(0)}$ with the
aid of charge conservation condition.
The second term describes the interaction via
transverse currents. It is especially important for relativistic
particles because $J_{c'c}/J^{(0)}_{c'c}\sim p_c/(m^*_c
c)$.
Longitudinal and transverse interactions are accompanied by
different plasma
screenings described by the functions $\Pi_t$ and $\Pi_l$
in the denominators of Eq.\ (\ref{Mfi}).

The collision energy and momentum transfers in neutron star cores
are typically small,  $\hbar\omega\sim k_B
T\ll \varepsilon_i$ and $\hbar q \ll p_{Fi}$. This smallness allows us to use the
weak-screening approximation which greatly simplifies the
consideration. Moreover, one typically has $\omega\ll qv_{Fi}$, so
that it is sufficient to use the asymptotic expressions (e.g., SY07)
\begin{eqnarray}
    \Pi_l&=&q_l^2=\frac{4\alpha}{\pi\hbar^2}\sum_i m_i^* p_{Fi}c, \\
    \Pi_t&=&i\frac{\pi}{4}\frac{\omega}{q c} q_{t}^2=i
    \frac{\alpha}{\hbar^2}\frac{\omega}{q c}\sum_i p_{Fi}^2,
\end{eqnarray}
where $\alpha=e^2/\hbar c \approx 1/137$ is the fine structure
constant; $q_l$ and $q_t$ are characteristic plasma
wavenumbers which depend on plasma composition (summation is
over all types of charged particles); $q_l$ is the familiar
Thomas-Fermi screening wavenumber; $q_t\lesssim q_l$,  with
$q_t\to q_l$ in the limit of ultra-relativistic particles.
Longitudinal interactions (via the exchange of longitudinal
plasmons) are mediated by static non-dissipative screening
with characteristic wavenumber $q_l$ ($\Pi_l$ is real), while
transverse interactions (via the exchange of
transverse plasmons) are accompanied by
the collisionless Landau damping
($\Pi_t$ is purely imaginary). Characteristic
momentum transfers in transverse interactions are
$\Lambda=(\pi\omega/(4cq_t))^{1/3}q_t \ll q_l$, meaning that
such interactions occur on larger spatial scales than
the longitudinal ones. Therefore, for relativistic
particles, the transverse interactions can be more efficient. The
importance of
such interactions
was pointed out by Heiselberg
and Pethick \cite{hp93} in their study of kinetic properties of
relativistic quark plasma. So far in all calculations of kinetic
properties in neutron star cores (except for SY07)
the transverse
interactions have been erroneously screened by the same static
dielectric function $\Pi_l$ as the longitudinal interactions.
This approximation strongly (up to
several orders of magnitude) overestimates the electron-muon thermal
conductivity (SY07). We will show that it overestimates also
$\eta_\mathrm{e\mu}$ (but less
dramatically).

The squared matrix element in (\ref{eq:Wci}) for
free ultra-relativistic particles can be written as
%
\begin{eqnarray}
   \langle |M_{ci}|^2\rangle&=&
   \frac{16\pi^2\hbar^6\alpha^2}{m_c^{*2}m_i^{*2}c^2}\varphi,
\\
   \frac{\langle |M_{cc}|^2\rangle}{2}&=&
   \frac{16\pi^2\hbar^6\alpha^2}{m_c^{*4}c^2}\left(\varphi-\gamma\right),
\end{eqnarray}
where $\varphi$ and $\gamma$ are dimensionless functions,
\begin{eqnarray}
   \varphi&=&\varphi_\parallel+\varphi_\perp+\varphi_{\perp\parallel},
\label{phi}\\
   \varphi_\parallel&=&\frac{(m_c^{*2}c^2-\hbar^2q^2/4)(m_i^{*2}c^2-\hbar^2q^2/4)}{\hbar^4(q^2+q^2_l)^2},
\label{phipar}\\
   \varphi_\perp&=&
   \frac{(p_{Fc}^2-\hbar^2q^2/4)
   (p_{Fi}^2-\hbar^2q^2/4)\cos^2\phi+\hbar^2(p_{Fc}^2+p_{Fi}^2)q^2/4}
   {\hbar^4(q^6+\Lambda^6)}\,q^2,
\label{phiper}\\
   \varphi_{\perp \parallel}&=&-2\,
   \frac{\sqrt{(p_{Fc}^2-\hbar^2q^2/4)(p_{Fi}^2-\hbar^2q^2/4)}}
   {\hbar^4(q^2+q^2_l)(q^6+\Lambda^6)}
   \, m_c^*m_i^*c^2q^4\cos\phi,
\label{phiparper}
\end{eqnarray}
$\phi$ being the angle between the vectors
$\bm{p}_{1}+\bm{p}_{1'}$ and $\bm{p}_{2}+\bm{p}_{2'}$. The
function $\gamma$ describes interference between two scattering
channels of identical particles. In the weak-screening
approximation, its contribution is small;
see Sec.\ \ref{corr}.

\subsection{Effective collision frequencies}
\label{relaxtimes}

The collision frequencies are obtained by calculating the
integrals (\ref{eq:nuci_int}) and (\ref{eq:nuci1_int}).
The calculations are greatly
simplified because all particles are strongly degenerate.
It is sufficient to
place the colliding particles on their Fermi surfaces
(whenever possible)
and use the standard energy-angular decomposition
based on ${\rm d}^3p=m^*\,p_F\,{\rm d} \varepsilon \,{\rm d} \Omega$,
where ${\rm d} \Omega$ is the solid angle element in the direction of
$\bm{p}$.
All (but one) energy
integrations can be done with the aid of the energy-conserving
delta-function in (\ref{eq:Wci}); only the $\omega$
integration is left. Three angular integrations out of eight
are performed
with the aid of the momentum-conserving delta-function; three
integrations  (over the position of $\bm{p}_1$
and over the azimuthal angle of $\bm{p}_2$ with respect to
$\bm{p}_1$) are trivial and give $8\pi^2$. As a result, one can reduce the
angular integration to the integration over ${\rm d} q$ and ${\rm
d} \phi$. Then the collision frequencies
(\ref{eq:nuci_int}) and (\ref{eq:nuci1_int}) can be written as
\begin{eqnarray}
  \nu_{ci}&=&\frac{12 \hbar^2\alpha^2}{\pi^2 p_{Fc}^5 m_c^* c^2}
  (k_B T)^2\int_0^\infty {\rm d} w\frac{w^2
  \exp(-w)}{\left[1-\exp(-w)\right]^2}
  I_{\Omega ci}(\omega),
\\
  \nu'_{ci}&=&\frac{12 \hbar^2\alpha^2}{\pi^2 p_{Fc}^5
  m_i^* c^2} (k_B T)^2\int_0^\infty {\rm d} w\frac{w^2
  \exp(-w)}{\left[1-\exp(-w)\right]^2} I'_{\Omega ci}(\omega),
\end{eqnarray}
where $w=\hbar\omega/(k_BT)$. The functions $I_{\Omega
ci}(\omega)$ and $I'_{\Omega ci}(\omega)$ are the angular
integrals
\begin{eqnarray}
   I_{\Omega ci}&=&\int_0^{q_m} {\rm d} q \int_0^\pi {\rm d} \phi\
   q^2\left(p_{Fc}^2-\frac{\hbar^2q^2}{4}\right) \varphi ,
\label{eq:intci}\\
   I_{\Omega ci}'&=&-\int_0^{q_m} {\rm d} q \int_0^\pi {\rm d} \phi\
   q^2\sqrt{(p_{Fc}^2-\hbar^2q^2/4)(p_{Fi}^2-\hbar^2q^2/4)}
   \cos\phi\  \varphi,
\label{intci1} 
\end{eqnarray}
where $\hbar q_m=\min\{2p_{c},2p_i\}$ is the maximum
momentum transfer in a collision event. Owing to a trivial
integration over $\phi$, $I_{\Omega ci}$ contains two terms coming
from 
$\varphi_\parallel$ and $\varphi_\perp$,
while $I_{\Omega ci}'$ contains only the
contribution from $\varphi_{\perp\parallel}$,
\begin{equation}
     I_{\Omega ci} = I_{\Omega ci}^{\parallel}+ I_{\Omega ci}^{\perp},
\ \ \ \ \ \ \
     I_{\Omega ci}'=I_{\Omega ci}^{\perp\parallel}.
\end{equation}

Let us calculate the angular integrals in the leading
approximation with respect to
the parameters $\Lambda/q_m$ and $q_l/q_m$. This approximation
is always justified for the transverse interactions because of the
presence of a small quantity $\hbar\omega\sim k_BT$ in the expression
for $\Lambda/q_m$. However, it is less accurate
for the longitudinal contribution since
$q_l/q_m$ is not too small; we will discuss corresponding
corrections  in
Sec.\ \ref{corr}.

The leading-order expressions for the angular
integrals are
\begin{eqnarray}
      I_{\Omega ci}^\perp&=&\frac{\pi^2}{6\hbar^4\Lambda}
      \,p_{Fc}^4p_{Fi}^2,
\label{intper}\\
     I^\parallel_{\Omega ci}&=&\frac{\pi^2}{4\hbar^4}
     \,\frac{m_c^{*2}m_i^{*2}p_{Fc}^2c^4}{q_l},
\label{intpar}\\
     I_{\Omega ci}^{\perp\parallel}&=&\frac{\pi^2 m_c^* m_i^*
      c^2p_{Fc}^2p_{Fi}^2}{2q_l\hbar^4}.
\label{intperpar}
\end{eqnarray}
Note that the leading-order expression for
$I_{\Omega ci}^{\parallel\perp}$ is independent of
$w$,
being of the
same order of magnitude
with respect to $q_l/q_m$ as $I_{\Omega ci}^{\parallel}$.
In contrast,
$I_{\Omega ci}^\perp\propto w^{-1/3} \propto \Lambda^{-1}$.

The final integration over $w$ gives the collision frequencies
\begin{eqnarray}
     \nu_{ci}^\perp&=&\frac{\xi\alpha^2}{\hbar^2 c}
     \;\frac{p_{Fi}^2}{p_{Fc}m_c^* c}
     \left(\frac{\hbar c}{q_t^2}\right)^{1/3} (k_BT)^{5/3},
\label{nuper} \\
     \nu_{ci}^{\parallel}&=&\frac{\pi^2\alpha^2
      m_c^*m_i^{*2}c^2}{\hbar^2 p_{Fc}^3 q_l} (k_BT)^2,
\label{nupar} \\
     \nu_{ci}'&=&\frac{2\pi^2\alpha^2m_c^*p_{Fi}^2}{\hbar^2 p_{Fc}^3
     q_l}(k_BT)^2,
\label{nucrossed}
\end{eqnarray}
where $\xi=2\Gamma(8/3)\zeta(5/3)(4/\pi)^{1/3}\approx 6.93$,
$\zeta(z)$ is the Riemann zeta function, and $\Gamma(z)$ is
the gamma function. Equations (\ref{eta}),
(\ref{taus})--(\ref{nomuons}), and (\ref{nuper})--(\ref{nucrossed})
give $\eta_\mathrm{e\mu}$
in the weak-screening approximation.

For typical conditions in a neutron star core,
\begin{equation}
    \nu_{ci}^\parallel\ll \nu_{ci}^\perp, \ \ \ \nu_{ci}' \ll \nu_{ci}^\perp.
\end{equation}
However, the inequality is not so strong as for the
thermal conductivity (SY07). The dominance of $\nu^\perp_{ci}$
over $\nu^\parallel_{ci}$
is determined by the factor $[\hbar
cq_l/(k_BT)]^{1/3}$ which increases slowly with decreasing
$T$. It is more accurate to include all components
of the collision frequencies.
For the thermal conductivity problem, we had
$\nu_{ci}^\perp/\nu_{ci}^\parallel \propto \hbar
cq_l/(k_BT)$, so that transverse interactions
dominated at all temperatures
of interest (SY07).

Nevertheless, for the not too high temperatures (see Sec.\
\ref{results} for details), $\eta_\mathrm{e\mu}$ is mainly determined
by the collisions via the exchange of transverse
plasmons. In this case, the electron and muon momentum transports
are decoupled [see Eq.\ (\ref{taus})],
\begin{equation}
\label{tauperp}
    \frac{1}{\tau_c}=\nu_c^\perp=\sum_i
     \nu_{ci}^\perp=
     \frac{\pi\xi}{4c^2}\;\frac{q_t}{p_{Fc}m_c^*}\,(\hbar c
     q_t)^{1/3} (k_BT)^{5/3}.
\end{equation}
Then the shear viscosity of electrons or muons
($c=$ e or $\mu$) becomes
\begin{equation}
\label{eq:etaperp}
    \eta_c=\eta_c^\perp=\frac{12\pi c^2\hbar^3}{5\xi} \;
    \frac{n_c^2}{q_t (\hbar c q_t)^{1/3}} \,(k_BT)^{-5/3}.
\end{equation}
We see, that in the low-temperature limit, $\eta_\mathrm{e\mu}$
has a non-standard temperature behavior, $\eta_\mathrm{e\mu}
\propto T^{-5/3}$ (instead of the standard Fermi-liquid dependence
$\eta\propto T^{-2}$). The non-standard behavior was pointed out
by Heiselberg and Pethick \cite{hp93} for an ultra-relativistic
quark plasma. Our results involve collisions of charged particles
in the $\mathrm{npe\mu}$-matter (for any degree of relativity of
muons). Our expressions for $\eta_\mathrm{e\mu}$ depend only on
the number densities of charged particles and on their effective
masses; therefore, they can be used for any EOS of dense matter.
Previous calculations \cite{fi79, fi76} overestimated
$\eta_{e\mu}$ because they employed the improper plasma screening
of transverse interactions. Equation (\ref{eq:etaperp}) remains
valid in the presence of other charged particles (such as
$\Sigma^-$ hyperons).

\subsection{Corrections to the leading terms}
\label{corr}

As mentioned above, the corrections to
$\nu_c^\perp$ containing
higher-order powers of $\Lambda/q_m$
can be
neglected, so that $\nu_c^\perp$
can be taken in the form (\ref{tauperp}). In contrast, the corrections to
$\nu_c^\parallel$ containing
higher-order powers of $q_l/q_m$ can be important. At
not too small temperatures, at which $\nu_c^\parallel$
can give a noticeable contribution,
such corrections can affect
$\eta_\mathrm{e\mu}$. We will discuss several
corrections of this type.

\subsubsection{Kinematical corrections to $\nu_{ci}^\parallel$
and $\nu_{ci}'$}
\label{corr_kin}

The main corrections to the leading terms arise from the
$q$-dependence of the functions $\varphi$ [Eqs.\
(\ref{phiper}) and (\ref{phiparper})] and
from the $q$-dependence in Eqs.\ (\ref{eq:intci})
and (\ref{intci1}).
The integral  $I^\parallel_{\Omega ci}$
is calculated precisely,
\begin{eqnarray}
\label{intparfull}
     I^\parallel_{\Omega ci}&=&
     \frac{\pi m_c^{*2}m_i^{*2}c^4p_{Fc}^2}{\hbar^4 q_l}\,
     I^\parallel_2(q_m/q_l)-\frac{\pi c^2 q_l}{4\hbar^2}
     \left[m_c^{2*}m_i^{*2} c^2
     +p_{Fc}^2(m_{c}^{*2}+m_i^{*2})\right]I^\parallel_4(q_m/q_l)\nonumber
\\
     &+&\frac{\pi q_l^3}{16}
     \left[(m_c^{*2}+m_i^{*2})c^2+p_{Fc}^2\right]
     I^\parallel_6(q_m/q_l)-\frac{\pi\hbar^2 q_l^5}{64}
     I^\parallel_8(q_m/q_l) ,
\end{eqnarray}
where we have introduced the integrals
\begin{equation}
   I^\parallel_{k}(x)=\int_0^x
   \frac{x'^k}{(x'^2+1)^2}\,{\rm d}x',
\label{E:parint}
\end{equation}
whose expressions are given in Appendix \ref{appendix}.
After the energy integration the corrected collision
frequency becomes
\begin{equation}
\label{nuparfull}
   \nu_{ci}^\parallel = \frac{4 \hbar^2\alpha^2}{ p_{Fc}^5 m_c^* c^2}
   (k_B T)^2 I_{\Omega ci}^\parallel.
\end{equation}

Similar corrections should be calculated for $\nu_{ci}'$. In the
leading-order approximation (with respect to $\Lambda/q_m$), $\Lambda^6$ can be
neglected in the denominator of $\varphi_{\perp\parallel}$.
The remaining angular integral is taken,
\begin{equation}
     I_{\Omega ci}^{\perp\parallel}=\frac{\pi m_i^* m_c^*c^2}{\hbar^4q_l}
     \left[p_{Fc}^2p_{Fi}^2
     I^{\perp\parallel}_0(q_m/q_l)-\frac{\hbar^2}{4}
     (p_{Fc}^2+p_{Fi}^2)q_l^2I^{\perp\parallel}_2(q_m/q_l)
     +\frac{\hbar^4}{16}q_l^4I^{\perp\parallel}_4(q_m/q_l)\right],
\end{equation}
where
\begin{equation}
   I^{\perp\parallel}_{k}(x)=\int_0^x
   \frac{x'^k}{x'^2+1}\,{\rm
   d}x'=I^{\parallel}_k(x)+I^{\parallel}_{k+2}(x).
\label{E:perparint}
\end{equation}
After the energy integration we finally obtain
\begin{equation}
    \nu'_{ci}=\frac{4 \hbar^2\alpha^2}{ p_{Fc}^5 m_i^* c^2} (k_B T)^2
    I_{\Omega ci}^{\perp\parallel}.
\end{equation}

Our calculations show that these corrections to $\nu_{ci}^\parallel$
and $\nu_{ci}'$ can reach
$\sim$70\%. It is advisable to include
them in $\eta_\mathrm{e\mu}$.

\subsubsection{Corrections to lepton-proton collision frequencies}

So far we have considered the function $\varphi$
calculated for free relativistic particles. It is a good
approximation for collisions within the electron-muon subsystem,
because the electrons and muons constitute almost ideal Fermi
gases. However, the protons belong to a strongly interacting Fermi
liquid; this case should be analyzed separately. First of all we
notice that the protons are non-relativistic. Moreover, we will
assume, that many-body effects can be treated by introducing an
effective proton mass $m^*_\mathrm{p}$. Under these assumptions, the proton
transition current can be written as $J_{p2'2}\propto
\frac{1}{2}(\bm{p}_2+\bm{p}_{2'})
m^*_\mathrm{p}\,\delta_{\sigma_2\sigma_{2'}}$, which only slightly modifies
$\varphi$. The expression (\ref{phiparper}) for
$\varphi_{\perp\parallel}$ remains the same, while the two other
functions become
\begin{eqnarray}
  \varphi^{c{\rm p}}_\parallel&=&\frac{(m_c^{*2}c^2-\hbar^2q^2/4)\,
  m_\mathrm{p}^{*2}c^2}{\hbar^4(q^2+q^2_l)^2},
\label{phicppar}\\
  \varphi^{c{\rm p}}_\perp&=&
  \frac{
  (p_{Fc}^2-\hbar^2 q^2/4)(p_{F \rm p}^2-\hbar^2 q^2/4)\cos^2\phi+
  (p_{Fc}^2-\hbar^2 q^2/4)\hbar^2q^2/4
  }{\hbar^4(q^6+\Lambda^6)}\,q^2.
\label{phicpper}
\end{eqnarray}
%

The difference
between $\nu_{c \rm p}^\parallel$, calculated with (\ref{phicppar}) and
(\ref{phipar}), is proportional to some
power of $\hbar q_l/(m_\mathrm{p}^*c)$.
Contrary to $q_l/q_m$, this ratio is always
small for the conditions in neutron star cores. Hence
Eq.\ (\ref{nuparfull}) with the angular integral
(\ref{intparfull}) remains a valid approximation.
For the completeness of our analysis, we
present the modified angular integral,
\begin{eqnarray}
\label{intparfull1}
   I^\parallel_{\Omega c{\rm p}}&=&\frac{\pi m_c^{*2}m_\mathrm{p}^{*2}c^4
   p_{Fc}^2}{\hbar^4 q_l} I^\parallel_2(q_m/q_l)-\frac{\pi c^2
   q_lm_{\rm p}^{*2}}{4\hbar^2}\left(m_c^{*2} c^2
   +p_{Fc}^2\right)I^\parallel_4(q_m/q_l)
\nonumber \\
   &+&\frac{\pi
   q_l^3}{16}\left(m_{\rm p}^{*2}c^2+p_{Fc}^2\right)I^\parallel_6(q_m/q_l).
\end{eqnarray}
This expression
gives
almost the same $\nu_{c{\rm p}}^\parallel$.

\subsubsection{Interference corrections to $\nu_{cc}$}

The last correction to be discussed is the correction to
$\nu_{cc}$ (to $\nu_\mathrm{ee}$ and $\nu_{\mu \mu}$)
due to the interference between two scattering channels ($1,2\to 1',
2'$ and $1,2 \to 2', 1'$) for collisions of identical particles.
This interference is described by the dimensionless function
$\gamma$ which (like $\varphi$) contains
longitudinal, transverse and mixed components.
An accurate consideration shows that all these components are smaller
than corresponding components of $\varphi$.
Therefore, the interference correction to
$\nu_{cc}^\perp$ can be neglected, and noticeable corrections
can arise only to
$\nu_{cc}^\parallel$ and $\nu_{cc}'$.  These
corrections have been calculated in the same way as in
previous sections. We have obtained that
they are numerically small (give $\lesssim5$\% contribution to
$\nu_{cc}^\parallel+\nu_{cc}'$). Their contribution
to $\eta_\mathrm{e\mu}$ is always
negligible as expected without
any numerical calculations. Such corrections
can be significant under two conditions.
First, the longitudinal component
$\nu_{cc}^\parallel$ should be comparable to $\nu_{cc}^\perp$. Second,
$\nu_{cc}$ itself should give a noticeable contribution to
$\eta_\mathrm{e\mu}$. The former condition would be realized
at high temperatures if particles
$c$ are weakly-relativistic. The electrons are ultra-relativistic
in neutron star cores and do not obey the above requirement.
The muons can be weakly-relativistic there, but if they are
their contribution to $\eta_\mathrm{e\mu}$ is not large.
The importance of the interference corrections in $cc$ collisions is
further reduced by a (typically) stronger contribution
of $c$p collisions. There are
also collisions between electrons and muons.
The interference corrections for such collisions are absent;
corresponding partial collision frequencies are of the
same order of magnitude as $\nu_{cc}$.

Thus, the corrections to $\nu_{c \rm p}$ and
$\nu_{cc}$ seem to be negligible.
The kinematical corrections to $\nu_{ci}^\parallel$
and $\nu_{ci}'$ are
significant if $\nu^\parallel_{ci}$ cannot be neglected in
comparison with $\nu^\perp_{ci}$. Note that in SY07, for the
electron-muon thermal conductivity problem, no corrections have
been required because the
thermal-conduction frequencies $\nu^\perp_{ci}$ dominate at
any density and temperature of practical interest.

\subsubsection{Comparison with exact solution}
\label{S:exact}

So far we have used a simplest variational solution for the shear viscosity based
on the expression for the trial function (\ref{Phi}) with
$\tau_{c}$ independent of the  particle energy $\varepsilon_c$.
Actually, however, the energy dependence of $\Phi_c$ is more complicated,
which affects the shear viscosity. It is convenient to introduce a
correction factor $C$ that relates the exact and variational
shear viscosities,
\begin{equation}
\label{eq:Cdef}
\eta_{\rm exact}=C\eta_{\rm var}.
\end{equation}
In ordinary Fermi-systems, where the collision probability is
independent of energy transfer $\hbar \omega$, the factor
$C$ can be calculated using the theory
developed by Sykes and Brooker \cite{sb70} for one component systems
and extended  by Flowers and Itoh \cite{fi79}
and Anderson et al.\ \cite{apq87} for multicomponent systems.
Unfortunately, this theory
cannot be directly applied to our case
because of the dynamical character of transverse plasma screening
(Landau damping).

The factor $C$ for the thermal conductivity with
account for the exchange of transverse plasmons was estimated in SY07.
Let us do the same for the shear viscosity. As in SY07, we
restrict ourselves to the exchange of transverse plasmons in
the weak-screening approximation. Then the electron transport
decouples from the muon one, and we can consider one type of
momentum carriers. We redefine $\Phi_c$ in (\ref{Phi}) as
\begin{equation}
\label{Phi_exact}
   \Phi_c=-\tau_{\rm eff} \left(v_{c\alpha} p_{c\beta} -\frac{1}{3}
   \delta_{\alpha\beta} v_cp_c\right){\rm V}_{\alpha \beta} \Psi(x),
\end{equation}
where $\tau_{\rm eff}$ is an effective relaxation time
(that can be treated as a normalization constant), and
an unknown function $\Psi(x)$ of $x=(\varepsilon_c-\mu_c)/(k_B T)$
describes the energy dependence of $\Phi_c$.

Substituting (\ref{Phi_exact}) into the linearized kinetic equation,
one obtains an integral equation for $\Psi(x)$,
\begin{eqnarray}
    f(x)(1-f(x))&=&\frac{6\hbar^4\alpha^2(k_BT)^2\tau_{\rm eff}}
    {\pi^2p_{Fc}^5 m_c^* c^2}\int\limits_{-\infty}^\infty {\rm
    d}x'\,
    \frac{x'-x}{\exp(x'-x)-1}f(x)(1-f(x'))\nonumber\\
&&\times
    \left\{ \frac{2}{3}\,\frac{p_{Fc}^4}{\hbar^2}I_{\Omega c1}
     (x'-x)[\Psi(x)-\Psi(x')] +I_{\Omega c}(x'-x) \Psi(x')\right\}
\label{inteq_exact},
\end{eqnarray}
where $I_{\Omega c}=\sum\limits_i I_{\Omega ci}$, $I_{\Omega
c1}=\sum\limits_i I_{\Omega ci1}$, and
\begin{equation}
   I_{\Omega ci 1}=\int_0^{q_m} {\rm d} q \int_0^\pi {\rm d} \phi\
   \varphi.\label{Ici1}
\end{equation}
The integral equation (\ref{inteq_exact}) is more complicated than
that for the thermal conductivity (see Eq.\ (42) in SY07). The
term with $I_{\Omega c1}$ in (\ref{inteq_exact})
appears because we go beyond the simplest
variational approach of $\Psi_c=const$.  Without that approach,
there is no cancellation of
zero-order expansion terms (in series of $q$) in kinematical factors in Eq.\
(\ref{eq:nuci_int}). It was that cancellation which led
to the appearance of the $q^2$ term  under the integral in
Eq.~(\ref{eq:intci}). The integral $I_{\Omega ci1}$ coincides
(save constant factor) with the angular integral for the thermal
conductivity problem (Eq.~(25) in SY07). Taking the weak-screening
expressions for $I_{\Omega c}$ and $I_{\Omega c1}$ with the exchange of
transverse plasmons alone, and choosing
\begin{equation}
   \tau_{\rm eff}=
   \left( 4 \over \pi \right)^{2/3}
   \frac{  p_{Fc}
   m_{c}^* c^2}{\alpha q_t(\hbar c q_t)^{1/3}
   (k_B T)^{5/3}}\label{taueff_exact},
\end{equation}
we obtain the dimensionless equation
\begin{eqnarray}
   \frac{1}{1+\exp(-x)}&=&\int\limits_{-\infty}^\infty {\rm d}x'\,
   \frac{\sgn(x'-x)}{(\exp(x'-x)-1)(1+\exp(-x'))}\nonumber\\
&&
    \times\left[ \lambda(\Psi(x)-\Psi(x')) +|x'-x|^{2/3}
    \Psi(x')\right],
\label{eqPsi_exact}
\end{eqnarray}
where $\lambda=p_{Fc}^2/(3\hbar^2\Lambda_T^2)$. The quantity
$\Lambda_T=(\pi k_BT/(4\hbar c q_t))^{1/3}q_t$ is the transverse
screening wavenumber $\Lambda$, with $\hbar\omega$ replaced
by $k_BT$. In a neutron star core,
one typically has $\hbar\Lambda_T\ll p_{Fc}$, and
hence $\lambda\gg 1$. If $\lambda\sim 1$, then the weak-screening
approximation is not justified.

Once $\Psi(x)$ is found by solving
Eq.~(\ref{eqPsi_exact}), the shear viscosity is given
by
\begin{equation}
   \eta_{c\rm exact}=\frac{n_c p_{Fc}^2 \tau_{\rm
   eff}}{5m_c^*}\int\limits_{-\infty}^{\infty}{\rm d} x\, \Psi(x)
   f(x)(1-f(x)).
\end{equation}
We have solved Eq.\ (\ref{eqPsi_exact}) numerically and compared the
result with the variational one ($\Psi_{\rm
var}=2^{2/3}\pi^{2/3}/[12\xi]$). For $\lambda=10-1000$,  we
obtain $C=\eta_{\rm exact}/\eta_{\rm var}=1.08-1.056$. For
$\lambda\geqslant 1000$, the factor $C=1.056$ becomes
nearly independent of $\lambda$.
Therefore, we have $C \approx 1$ indicating that
the simplest variational approach is well justified.

Were the electron (and muon) collisions determined solely by the
exchange of longitudinal plasmons (with the transition matrix element
independent of $\omega$), one could find $C$ from the standard theory.
In that case one also obtains $C \approx 1$
(see, e.g., Ref.~\cite{bh99}). We expect, that in the
most general case, when electron and muon collisions are
governed by the exchange of transverse and longitudinal plasmons, the
correction factor $C$ differs from $C=1$ by $\lesssim 10\%$.
If so, the simplest variational approach is sufficiently accurate,
and no corrections are required ($C=1$)
for the majority of astrophysical applications.


\subsection{Neutron viscosity}
\label{viscn_norm}

In this section we calculate the neutron shear viscosity
$\eta_\mathrm{n}$. We employ the same formalism as was used
in BHY01 for studying the neutron thermal
conductivity. Similar approach was used by Baiko and Haensel
\cite{bh99} to determine kinetic coefficients
mediated by neutron-neutron collisions. We calculate
$\eta_\mathrm{nn}$ from Eq.\ (\ref{eta}); the effective relaxation
time of neutrons, $\tau_\mathrm{n}$, is given by Eq.\ (\ref{taus}),
being determined by the collision frequencies (\ref{nus}) of
neutrons with neutrons and protons.

The neutron-neutron collision frequency can
be written as
\begin{equation}
\label{E:nunnFinal}
   \nu_\mathrm{nn}+\nu_\mathrm{nn}'=
   \frac{16m_\mathrm{n}^{*3}(k_BT)^2 }{3m_\mathrm{n}^2\hbar^3}
   S_\mathrm{nn}
\end{equation}
(note that the authors of BHY01 did not
separate
$\nu_\mathrm{nn}$ and $\nu_\mathrm{nn}'$ but considered their sum).
The neutron-proton collision frequency
is
\begin{equation}
\label{E:nunp}
   \nu_\mathrm{np}=
   \frac{32m_\mathrm{p}^{*2}m_\mathrm{n}^*(k_BT)^2}{3m_\mathrm{n}^2\hbar^3}
   S_\mathrm{np}.
\end{equation}
Here, $m_\mathrm{n}$
is a bare nucleon mass. The quantities
$S_\mathrm{nn}$ and $S_\mathrm{np}$ are the effective nucleon-nucleon
scattering cross-sections
introduced in Eq.\ (22) of BHY01 (for the thermal
conduction problem). For the shear
viscosity, in the same notations as in BHY01, we obtain
\begin{eqnarray}
    S_\mathrm{nn}&=&\frac{m_\mathrm{n}^2}{16\pi^2\hbar^4}
    \int_0^1 {\rm d}x'\,\int_0^{\sqrt{1-x'^2}} {\rm d}x\,
   \frac{12x^2x'^2}{\sqrt{1-x^2-x'^2}} \;{\cal Q}_\mathrm{nn},
\label{E:Snn}\\
   S_\mathrm{np}&=&\frac{m_\mathrm{n}^2}{16\pi^2\hbar^4}
   \int_{0.5-x_0}^{0.5+x_0} {\rm d}x'\,\int_0^{a}{\rm d}x\,
   \frac{6(x^2-x^4)}{\sqrt{a^2-x^2}}\;{\cal Q}_\mathrm{np},
\label{E:Snp}
\end{eqnarray}
where $x=\hbar q/(2p_{F\rm n})$, $x'=\hbar q'/(2p_{F\rm n})$,
$a=\sqrt{x_0^2-(0.25+x_0^2-x'^2)^2}/x'$,
and $x_0=p_{F \rm p}/(2p_{F\rm n})$.
This choice of integration variables is
convenient for numerical integration. The
quantities ${\cal Q}_\mathrm{nn}$
and ${\cal Q}_\mathrm{np}$ are
squared matrix elements for nucleon-nucleon scattering
(in the notations of BHY01,
${\cal Q}_\mathrm{nn}=\langle|M_\mathrm{nn}|^2\rangle$ and
${\cal Q}_\mathrm{np}=\langle|M_\mathrm{np}|^2\rangle$).

Let us emphasize that kinematic restrictions
in Eqs.\ (\ref{E:Snn}) and (\ref{E:Snp}) are very different.
The effective cross section $S_\mathrm{nn}$ is
determined by a wide
spectrum of momentum transfers $q$ (or, equivalently,
of scattering angles).
Our calculations show that one can get a reasonably
accurate $S_\mathrm{nn}$ assuming that ${\cal Q}_\mathrm{nn}$
is independent of $q$. Such a $q$-averaged ${\cal Q}_\mathrm{nn}$
can be extracted from a total neutron-neutron scattering cross
section. In contrast (because, typically, $p_{F \rm p} \ll p_{F \rm n}$),
$S_\mathrm{np}$ is determined by small momentum transfers, that is
by a small-angle cross section of neutron-proton scattering.

By analogy with BHY01, we can write $S_{\rm nn}=S^{(0)}_{\rm nn}
K_{\rm nn}$ and $S_{\rm np}=S^{(0)}_{\rm np} K_{\rm np}$. Here,
$S_{\rm nn}$ and $S_{\rm np}$ are the effective cross sections
calculated with in-medium squared matrix elements, ${\cal
Q}_\mathrm{nn}$ and ${\cal Q}_\mathrm{np}$; $S_{\rm nn}^{(0)}$ and
$S_{\rm np}^{(0)}$ are similar cross sections calculated with the
in-vacuum matrix elements; $K_{\rm nn}$ and $K_{\rm np}$ are the
ratios of the in-medium to the in-vacuum cross sections.

The authors of BHY01 calculated
all these quantities for the thermal conduction problem.
The squared matrix elements
${\cal Q}_\mathrm{nn}$ and ${\cal Q}_\mathrm{np}$ were
extracted from nucleon-nucleon differential
scattering cross sections  calculated in Refs.\ \cite{LM93,LM94} for
symmetric nuclear matter with the Bonn nucleon-nucleon
interaction potential using the Dirac-Brueckner approach.
An accurate extraction of the in-medium ${\cal Q}_\mathrm{nn}$ and
${\cal Q}_\mathrm{np}$ required the knowledge of effective masses
$m_{\rm n}^*$ and $m_{\rm p}^*$
(not reported in \cite{LM93,LM94}). For that reason,
the procedure used in BHY01 was ambiguous. Thus,
the factors $K_{\rm nn}$ and
$K_{\rm np}$, presented in BHY01 for the neutron
thermal conductivity, are model dependent
and not very certain.

Now we turn to calculating $S_{\rm nn}$, $S_{\rm np}$, $S_{\rm
nn}^{(0)}$, $S_{\rm np}^{(0)}$, $K_{\rm nn}$, and $K_{\rm np}$ for
the shear viscosity. To avoid the above drawbacks, we suggest to
neglect the in-medium effects on the squared matrix elements and
set $K_{\rm nn}=K_{\rm np}=1$, $S_{\rm nn}=S_{\rm nn}^{(0)}$, and
$S_{\rm np}=S_{\rm np}^{(0)}$. According to BHY01 (for the thermal
conductivity), $K_{\rm nn}$ and $K_{\rm np}$ are indeed $\sim 1$
(and $K_{\rm np}$ is relatively unimportant). Recently Zhang et
al.\ \cite{zhangetal07} have studied nucleon-nucleon scattering
cross sections in nuclear matter taking into account two-nucleon
and three-nucleon interactions. They used the
Brueckner-Hartree-Fock approach and the Argonne V14
nucleon-nucleon interaction model supplemented by three-nucleon
interactions. Their principal conclusion is that the in-medium
effects on square matrix elements are relatively weak, while the
main medium effect consists in modifying (mostly reducing) $m_{\rm
n}^*$ and $m_{\rm p}^*$. The reduction of effective masses under
the simultaneous effects of two-nucleon and three-nucleon forces
is much stronger than under the effect of two-nucleon forces
alone.

Thus, we have calculated ${\cal S}_\mathrm{nn}$ and ${\cal
S}_\mathrm{np}$ from Eqs.\ (\ref{E:Snn}) and (\ref{E:Snp}) using
the in-vacuum matrix elements ${\cal Q}_\mathrm{nn}$ and ${\cal
Q}_\mathrm{np}$ from Refs.\ \cite{LM93,LM94}. These matrix
elements accurately reproduce \cite{mhe87} well elaborated laboratory
measurements of differential nucleon-nucleon scattering cross
sections. Our calculations of ${\cal S}_\mathrm{nn}$
and ${\cal S}_\mathrm{np}$ are expected to be very close to those
done with in-vacuum cross sections measured in laboratory. In this
sense, our values of ${\cal S}_\mathrm{nn}$ and ${\cal
S}_\mathrm{np}$ are model independent. Similar calculations of
${\cal S}_\mathrm{nn}$ in Ref.\ \cite{bh99} give slightly
different results due to the different data sets for  ${\cal
Q}_\mathrm{nn}$. Because our equations (\ref{E:nunnFinal}) and
(\ref{E:nunp}) for the nucleon-nucleon collision frequencies
contain a proper dependence on $m_{\rm n}^*$ and $m_{\rm p}^*$,
they can be regarded as {\it
independent} of any specific model for nucleon-nucleon
interaction. It can be a two-body or two-body plus three-body
interaction; its explicit form is not essential. Thus, we obtain a
description of the neutron shear viscosity valid for any EOS of
nucleon matter. This approach is not strict (uses the in-vacuum
matrix elements) but universal. One can in principle calculate
more accurate in-medium matrix elements for any chosen EOS but
loosing the universality. For calculating the diffusive thermal
conductivity from the equations of BHY01, we would recommend to
adopt the same approach and set $K_{\rm nn}=K_{\rm np}=1$ in those
equations.

The results of our calculations can be fitted by the expressions
\begin{eqnarray}
   S_\mathrm{nn}&=&\frac{12.88}{k_\mathrm{n}^{1.915}}\;
             \frac{1-0.6253k_\mathrm{n}+0.3305k_\mathrm{n}^2}{
         1-0.0736k_\mathrm{n}}~~{\rm mb},
\nonumber \\
   S_\mathrm{np}&=&\frac{0.8876\,
             k_\mathrm{p}^{3.5}}{k_\mathrm{n}^5}\;
             \frac{1+139.6k_{\rm p}+103.7k_\mathrm{n}}
         {1-0.5932k_\mathrm{n}+0.1829k_\mathrm{n}^2
     +7.629k_\mathrm{p}^2-0.5405k_\mathrm{p}k_\mathrm{n}}~~{\rm mb},
\label{E:Sfits}
\end{eqnarray}
%
%
where $k_i$ is the Fermi wavenumber of nucleons $i$ expressed in
fm$^{-1}$. As in BHY01, the calculations and fits cover the range
of $k_\mathrm{n}$ from 1.1 to 2.6 fm$^{-1}$ and the range of
$k_\mathrm{p}$ from 0.3 to 1.2 fm$^{-1}$. These parameter ranges
are appropriate to neutron star cores at $0.5 \, \rho_0 \lesssim
\rho \lesssim 3 \rho_0$. The fit errors for $S_\mathrm{nn}$ do not
exceed $0.5\%$. The maximum fit error of $S_\mathrm{np}$ is
$\delta_{\rm max} \sim 8\%$ (at $k_{\rm n}=1.1$ fm$^{-1}$ and
$k_{\rm p}=0.7$ fm$^{-1}$).


\section{Shear viscosity in superfluid matter}
\label{superflu}

Neutrons and protons in the $\mathrm{npe\mu}$-matter of neutron
star cores can be in superfluid state (e.g., Ref.\ \cite{ls01}).
Here, we study the effects
of superfluidity on the shear viscosity. Let $T_\mathrm{c \rm n}(\rho)$
be the critical temperature for superfluidity of neutrons, and
$T_\mathrm{c \rm p}(\rho)$ be the same for protons. Proton
superfluidity means superconductivity. Calculations of
superfluid critical temperatures are complicated and very
sensitive to a chosen model of nucleon-nucleon interaction and a
method to employ many-body (polarization) effects \cite{ls01}. Numerous
calculations give drastically different $T_\mathrm{c\rm n}(\rho)$ and
$T_\mathrm{c\rm p}(\rho)$. It is instructive not to rely
on any particular model but treat $T_\mathrm{cn}$ and
$T_\mathrm{c \rm p}$ as free parameters varied within reasonable limits
($T_\mathrm{c} \lesssim 10^{10}$~K), in accordance with
microscopic calculations.

Neutron superfluidity has no direct effect on the
shear viscosity $\eta_\mathrm{e\mu}$
of electrons and muons (because $\eta_\mathrm{e\mu}$
is limited by electromagnetic interactions). However, it strongly
affects neutron star hydrodynamics in a complicated way.
It makes the hydrodynamics essentially multifluid
(with several hydrodynamical velocity fields;
e.g., \cite{gusakov07} and references therein); it introduces
an entrainment effect (which relates motion of neutrons
and protons), creates a very specific spectrum of
elementary medium excitations (phonons) and associated
specific energy and momentum transfer mechanisms.
All these problems go far beyond
the scope of our paper. Therefore, we will neglect the effects of
neutron superfluidity (will treat neutrons as normal)
but consider the effects of proton superfluidity on $\eta_{\rm n}$
(assuming the protons to be passive scatterers of neutrons, i.e.,
ignoring momentum transport by protons). We disregard thus
hydrodynamical effects of neutron and proton
superfluids.

Microscopically, proton superfluidity manifests itself
in rearranging proton states (from normal Fermi-liquid
quasiparticles to Bogoliubov quasiparticles)
and in the appearance of a gap $\Delta$
in the proton energy spectrum near the Fermi level
($\varepsilon=\mu$),
\begin{equation}
\label{E:DispersSF}
    \varepsilon=\mu+\sgn(\xi)\sqrt{\Delta^2+\xi^2},
\end{equation}
where $\xi\equiv v_{F}(p-p_{ F})$; the presented equation is valid
at $|\xi| \ll \mu$.

It is generally believed, that Cooper paring of protons appears in the
singlet $^1$S$_0$ state (e.g., Ref.\ \cite{ls01}).
The temperature dependence
of $\Delta$, calculated in the BCS approximation, can be
approximated as \cite{LYheat} 
\begin{equation}
\label{E:delta0}
     y=\frac{\Delta}{k_B T}=
     \sqrt{1-\tau}
     \left(1.456-\frac{0.157}{\sqrt{\tau}}+\frac{1.764}{\tau}\right),
\end{equation}
where $\tau=T/T_\mathrm{cp}$.

\subsection{The effects of proton superfluidity on
the electron-muon viscosity}
\label{eta_emu_sf}

The effects of proton superfluidity on $\eta_\mathrm{e\mu}$ are
twofold. First, proton superfluidity affects the plasma dielectric
function and, hence, the screening of electromagnetic
interactions. The longitudinal dielectric function is almost
insensitive to the presence of superfluidity \cite{alf06}, while
the transverse dielectric function modifies  collision frequencies
$\nu^\perp_{ci}$. The frequencies $\nu'_{ci}$ remain almost
unchanged because they are independent of the transverse screening
in the leading order.

As in SY07, a collision frequency
in superfluid matter (to be denoted as  $\nu^{\perp S}_{ci}$)
can be written as
\begin{equation}
   \nu_{ci}^{\perp S}=\nu_{ci}^{\perp} R_l^\perp(y,r),
\end{equation}
where $\nu^\perp_{ci}$ ($i=$e, $\mu$)
stands for a collision frequency
in non-superfluid matter, while $R^\perp_l(y,r)$
accounts for the superfluid effects
(which mainly reduce the collision rate);
\begin{equation}
   r=(p_{F \rm e}^2+p_{F\mu}^2)/p_{F \rm p}^2
\end{equation}
is a slowly varying function determined by plasma composition. We
have $r=1$ in the absence of muons, and $r>1$ in the presence of
muons, with the maximum value of
$r\approx 1.26$ in the limit of ultra-relativistic muons.
The reduction factor $R_l^\perp(y,r)$
for the shear viscosity is, however, not the same as for the thermal
conductivity (obtained in SY07) and will be calculated below.

The second effect of proton superfluidity consists in an
additional (direct; not through plasma screening)
reduction of the lepton-proton
collision frequencies
(such as $\nu_\mathrm{ep}$ and $\nu_\mathrm{\mu p}$).
This direct reduction is exponential; it can be described by the
reduction factors
\begin{eqnarray}
   \nu_{c \rm p}^{\perp S}&=&\nu_{c \rm p}^{\perp }
   R_\mathrm{p}^\perp(y,r),
\\
   \nu_{c \rm p}^{\parallel S}&=&\nu_{c \rm p}^{\parallel }
   R_\mathrm{p}^\parallel(y).
\end{eqnarray}
The reduction factors $R_\mathrm{p}^\perp$ and $R_\mathrm{p}^\parallel$ are not
the same due to the difference in longitudinal and transverse
plasma screenings. Following SY07, it is convenient to introduce the
reduction factor $R_{tot}^\perp$ for the total transverse collision frequency
$\nu_c^\perp =\nu_{c\rm e}^\perp+\nu_{c\mu}^\perp+\nu_{c \rm
p}^\perp$,
\begin{eqnarray}
    \nu_c^{\perp S}&=&\nu_c^{\perp S}R_{tot}^\perp(y,r),
\\
    R_{tot}^\perp(y,r)&=&
    \left[rR_l^\perp(y,r)+R_{\rm p}^\perp(y,r)\right]/(r+1).
\end{eqnarray}
Below we calculate
$R_l^\perp$, $R_{\rm p}^\parallel$, $R_{\rm p}^\perp$, and $R_{tot}^\perp$.

\subsubsection{Superfluid reduction of collisions
in electron-muon subsystem}
\label{nu_emu_superfl}

Superfluid reduction of lepton-lepton collisions is governed by
the transverse polarization function $\Pi_t$. For the conditions
in neutron star cores, it is sufficient to use $\Pi_t$ in the
so-called Pippard limit ($\hbar\omega \ll p_{F \rm p}v_{F \rm p}$,
$\hbar q \ll p_{F \rm p}$
 and $\xi\gg 1/q$, where $\xi \sim
\hbar v_{F \rm p}/(k_B T_{c \rm p})$ is the coherence length). In this
approximation, the proton contribution to $\Pi_t$ reads 
\begin{equation}
  \Pi_t^{\rm (p)}=\frac{q_{t \rm p}^2}{4}\,\frac{\Delta}{\hbar cq}
  \,Q(w,y),
\end{equation}
where $q_{ti}^2=4\alpha p_{Fi}^2/(\hbar^2 \pi)$, and
$Q$ is the response function calculated
in Ref.\ 
\cite{Mattis58} and discussed in
SY07 in more details.

In the non-superfluid limit of $y\ll 1$ one has $Q=i\pi \hbar
w/y$, which corresponds to the standard Landau-damping expression.
In the opposite case of strong superfluidity ($y \gg 1$), the
response function $Q$ becomes pure real, $Q=\pi^2$. For
intermediate superfluidity, $y \sim 1$, we have used the
expressions for $Q$ derived in \cite{Mattis58}. They are valid for
a pure BCS formalism neglecting collective modes and related
vortex renormalization in current operators due to gradient
invariance. However, as in SY07, the main contribution to
$\eta_\mathrm{e\mu}$ comes from the parameter values far from
characteristic frequencies  of collective modes (far from
$\omega\sim v_Fq$) and we can use the standard BCS theory.

The expression for the total polarization function in the
superfluid case takes the form
\begin{equation}
  \Pi_{t}=\frac{\pi\omega}{4qc}
  \left\{ q_{t \rm p}^2\,\frac{y}{\pi w}
  \,\Re
  Q(w,y) +i\left[q_{t\rm e}^2+q_{t\mu}^2 +
  q_{t \rm p}^2\,\frac{y}{\pi w}\,\Im Q(w,y)
  \right] \right\}.
\end{equation}
%
In the case of strong superfluidity, the main contribution to
$\Pi_{t}$ comes from protons. Moreover, the character of plasma
screening changes. Instead of the dynamical Landau damping, the
screening becomes static, with the frequency-independent screening
wave number $\Lambda_S=[\pi^2q_{t\rm p}^2\Delta/(4\hbar
c)]^{1/3}$. In neutron star cores, one typically has $\Delta\sim
k_BT_{c\rm p}\ll p_{Fi}c$. Therefore, the relation $\Lambda_S \ll
q_l$ remains true in the superfluid case. In other words, the
exchange of transverse plasmons in proton superfluid remains more
efficient than the exchange of longitudinal plasmons. The strong
inequality $\Lambda_S \ll q_m$ justifies the use of the
leading-order weak screening approximation in describing the
exchange of transverse plasmons.

In order to calculate the reduction factor $R_l^\perp$ one should
reconsider the transverse angular integral $I_{\Omega ci}^\perp$
taking into account the changes of electrodynamical plasma properties
in superfluid matter.
In the leading order with respect to
$\Lambda_S/q_m$,
\begin{eqnarray}
  I_{\Omega ci}^{\perp S} &=& I_{\Omega ci}^{\perp}\,
  F^\perp(w,y,r),
\end{eqnarray}
where $I_{\Omega ci}^{\perp}$ refers to non-superfluid matter,
while
\begin{eqnarray}
   F^{\perp}(w,y,r)&=&
   \frac{\left[\pi w(r+1)\right]^{1/3}
   \left[\left(\pi w
   r+y\Im Q(w,y)\right)^2+ \left(y\Re
   Q(w,y)\right)^2\right]^{1/3}}{|\pi w r+\Delta\Im
   Q(w,y)|}
\nonumber\\
    &\times&\frac{2}{\sqrt{3}}
    \sin\left[\frac{2}{3}\arctan\frac{|\pi w
   r+y\Im Q(w,y)|}{y\Re\, Q(w,y)}
\right]
\end{eqnarray}
accounts for superfluid effects. In the limit of
strong superfluidity ($y\gg 1$) we have
\begin{eqnarray}
   F^{\perp}(w,y,r)&=&\frac{4}{3\sqrt{3}}
   \left[\frac{w(r+1)}{\pi y}\right]^{1/3}.
\end{eqnarray}
This asymptotic $w$-dependence compensates the $w$-dependence in
$I_{\Omega ci}^{\perp S}$ (that appeared under the effect of
plasma screening). Moreover, in the expression for $I_{\Omega
ci}^{\perp S}$ the collision energy $\hbar \omega$ is now replaced
by the energy gap $\Delta$.

Finally, we write  $\nu_{ci}^{\perp S}= \nu_{ci}^{\perp } R_l^\perp(y,r) $,
and the reduction factor becomes
\begin{eqnarray}
  R^{\perp}_l(y,r)&=&
  \frac{1}{\Gamma(8/3)\zeta(5/3)}\int_0^\infty
  {\exp(w) \over [\exp(w)-1]^2 }\, w^{5/3}  F^{\perp}(w,y,r)\, {\rm d}w.
\end{eqnarray}
%
When superfluidity vanishes ($y\to 1$) we evidently
have $R^\perp_l(y,r)\to 1$.
In the opposite case of strong superfluidity  ($y\gg 1$) we
obtain
\begin{eqnarray}
  R^{\perp}_l(y,r)&=&\frac{4\pi^2}{9\sqrt{3}\Gamma(8/3)\zeta(5/3)}
  \left(\frac{r+1}{\pi y}\right)^{1/3}.
\end{eqnarray}
Thus, strong proton superfluidity restores the temperature
dependence $\nu_{ci}^{\perp S}\propto T^2$ that is standard for
Fermi systems. This result was derived in SY07 for the
thermal conduction problem. It is a natural
consequence of changing plasma screening from
dynamical to statical one when $T$ falls below $T_\mathrm{c \rm p}$.

In addition, we have
computed $R^\perp_l(y,r)$ for a wide grid of $y$. We do not
present an appropriate fit, because we will
calculate and fit the total reduction factor
$R_{tot}^\perp(y,r)$ for $\nu_c^\perp$.

\subsubsection{Superfluid reduction of collisions of electrons
and muons with protons}
\label{red_prot}

Now consider a direct effect of superfluidity
on electron-proton and muon-proton collision rates.
The consideration is similar to that for the thermal
conduction problem (SY07). The proton energy gap
has to be included  in the expressions for the collision frequencies
through the proton Fermi-Dirac
distributions.
In addition, the electron-proton and muon-proton
scattering matrix elements have to be calculated
using wave functions of proton Bogoliubov quasiparticles.
As a result, the
reduction factors $R^\perp_{\rm p}(y)$ and $R^\parallel_{\rm p}(y,r)$ can be
written as
\begin{eqnarray}
 R^\perp_{\rm p}(y,r)&=&\frac{1}{\Gamma(8/3)\zeta(5/3)}
     \int\limits_0^\infty\int\limits_0^\infty
     \frac{{\rm d}x_2\,{\rm d}x_{2'}}{1+\exp(z_2)}\nonumber\\
&\times&
    \left\{\frac{(z_{2'}-z_2)|z_{2'}-z_2|^{-1/3}
    \left(1+4\rm{u}_2\rm{u}_{2'}\rm{v}_2\rm{v}_{2'}\right)}
    {[1+\exp(-z_{2'})][\exp(z_{2'}-z_2)-1]}
    F^\perp(|z_{2'}-z_2|,y,r)\right.\nonumber\\
&-&\left. \frac{(z_{2'}+z_2)|z_{2'}+z_2|^{-1/3}
     \left(1-4 \rm{u}_2\rm{u}_{2'}\rm{v}_2 \rm{v}_{2'}\right)}
     {[1+\exp(z_{2'})][\exp(-z_{2'}-z_2)-1]}
     F^\perp(|z_{2'}+z_2|,y,r)
\right\},\\
R^\parallel_{\rm p}(y)&=&
     \frac{3}{\pi^2}\int\limits_0^\infty\int\limits_0^\infty
     \frac{{\rm d}x_2\,{\rm d}x_{2'}}{1+\exp(z_2)}
     \left\{\frac{(z_{2'}-z_2)(1-4\rm{u}_2\rm{u}_{2'}\rm{v}_2\rm{v}_{2'})}
     {[1+\exp(-z_{2'})][\exp(z_{2'}-z_2)-1]} \right.\nonumber\\
&-&\left.
     \frac{(z_{2'}+z_2)(1+4\rm{u}_2\rm{u}_{2'}{\rm v}_2{\rm v}_{2'})}
     {[1+\exp(z_{2'})][\exp(-z_{2'}-z_2)-1]}
     \right\},
\end{eqnarray}
where
\begin{eqnarray}
  \mathrm{u}_{\rm p}&=&\frac{1}{\sqrt{2}}\,\sqrt{1+\frac{x}{z}},
\nonumber  \\
  \mathrm{v}_{\rm p}&=&\frac{\sgn(x)}{\sqrt{2}}\,\sqrt{1-\frac{x}{z}},
\label{uv}
\end{eqnarray}
$x=v_{F \rm p}(p-p_{F \rm p})/(k_BT)$ and
$z=(\varepsilon-\mu_{\rm p})/(k_BT)$.

In the limit of strong superfluidity ($y\gg 1$) we obtain
$R^\parallel_{\rm p}(y)=A^\parallel \exp(-y)$
and $R^\perp_{\rm p}(y,r)=A^\perp (r+1)^{1/3} y^{2/3} \exp(-y)$, where
\begin{equation}
A^\parallel=\frac{6}{\pi^2}\int_0^\infty {\rm d}\eta_1
\int_0^\infty
   {\rm d}\eta_2 \,
   \frac{(\eta_1^2-\eta_2^2)(\eta_1^2+\eta_2^2)}
   {\exp(\eta_1^2)-\exp(\eta_2^2)}\approx
   1.45425
\end{equation}
and
\begin{equation}
   A^\perp=\frac{16}{3\sqrt{3}\pi^{1/3}\Gamma(8/3)\zeta(5/3)}\int_0^\infty
   {\rm d}\eta_1 \int_0^\infty {\rm d}\eta_2 \,
   \frac{\eta_1^2-\eta_2^2}{\exp(\eta_1^2)-\exp(\eta_2^2)}\approx
   0.92974.
\end{equation}
Thus, at $T \ll T_\mathrm{cp}$  collisions with superfluid protons are
exponentially suppressed. Then the
shear viscosity $\eta_\mathrm{e \mu}$ is limited
by collisions within the
electron-muon subsystem (which are also affected by
proton superfluidity as described in Sec.\  \ref{nu_emu_superfl}).

We have computed $R^\parallel_{\rm p}(y)$ for a wide range of
$y$ and fitted the results by the expression
\begin{eqnarray}
    R^\parallel_{\rm p}(y)&=&\left\{A^\parallel+
    (1.25-A^\parallel)\exp(-0.0437\;y) \right.
\nonumber \\
    &+& \left. (1.473\;y^2+0.00618\;y^4)
    \exp\left[0.42-\sqrt{(0.42)^2+y^2}\right]\right\}
\nonumber\\
    &\times&
    \exp\left[-\sqrt{(0.22)^2+y^2}\right],
\label{E:Rparsfit}
\end{eqnarray}
which reproduces also the asymptotic limits.
The maximum relative fit error is $0.75\%$ at $y=0.533$.

We do not present a separate fit expression for
$R^\perp_{\rm p}(y,r)$, but give the fit of the total
reduction factor $R^\perp_{tot}(y,r)$:
\begin{eqnarray}
  R_{tot}^\perp&=&\frac{1-g_1}{\left(1+g_3y^3\right)^{1/9}}
  +\left(g_1+g_2\right)\exp\left[0.145-\sqrt{(0.145)^2+y^2}\right],
\\
  g_1&=&0.87-0.314r,\ \ \ \
  g_2=(0.423+0.003r)y^{1/3}+0.0146y^2-0.598y^{1/3}\exp(-y),
\nonumber  \\
  g_3&=&251 r^{-9}(r+1)^6(1-g_1)^9,
\nonumber
\end{eqnarray}
with the maximum fit error $\sim 0.3\%$ at $r=1$ and $y=3.5$. This fit
reproduces also the limiting case of $R^\perp_{tot}\to 1$ at $y\to 0$,
and the asymptote at $y \gg 1$,
\begin{eqnarray}
  R^{\perp}_{tot}(y,r)&=&
  \frac{4\pi^2r}{9\sqrt{3}\Gamma(8/3)\zeta(5/3)(r+1)^{2/3}}\;
  \frac{1}{(\pi y)^{1/3}}.
\end{eqnarray}

Recently the electron shear viscosity in superfluid matter has been
analyzed by Andersson et al.\ \cite{acg05}. These authors have
used the standard  (but approximate)  approach
in which the transverse plasma screening
is assumed to be the same
as the longitudinal one.
This approach is inaccurate
even in the non-superfluid case. In Ref.\
\cite{acg05} the
effects of superfluidity are described by a reduction factor
$R^*_\textrm{ep}$
for the effective electron-proton
collision frequency. That
factor has been taken from Ref.\ \cite{gy95}
devoted to the thermal conductivity problem. However, the
reduction factors for the thermal conductivity and shear viscosity
are different. Moreover, the factor
$R^*_\textrm{ep}$ in  \cite{acg05}  is
inaccurate even for the thermal conductivity, because
it assumes approximate plasma screening
and
neglects additional terms associated with
creation/annihilation of proton Bogoliubov quasiparticles; see SY07 for
details. Nevertheless, numerical values of $\eta_\mathrm{e}$,
derived from the results of \cite{acg05} for superfluid
matter, are not too different from our results.
Typically, they overestimate $\eta_\mathrm{e}$
by a factor of three, and this overestimation increases with
decreasing $T_{c \rm p}$.

In the limit of strong superfluidity, the temperature dependence of
$\nu_{c}^{\perp S}$ formally restores the standard Fermi-liquid
behavior, $\nu_c^{\perp S}\propto T^2$. Therefore,
$\nu_{ci}^\parallel$ can be comparable to $\nu_{c}^{\perp S}$.
Note, that the ratio
$\nu_{ci}^\parallel/\nu_{c}^{\perp S}$
in superfluid matter remains approximately the same as
its value  at
$T=T_{c \rm p}$. If $T_{c \rm p}$ is
sufficiently small, then at $T=T_\mathrm{cp}$
we have $\nu_{c}^{\perp} \gg \nu_{ci}^\parallel$,
and the same inequality holds at smaller $T$.
The shear viscosity in superfluid
matter, fully determined by the exchange
of transverse plasmons, can be
written as
\begin{equation}
\label{etaper_sf}
  \eta_c^{ \perp S}={\xi_S \over (k_BT)^2} \, \frac{n_c^2}
  {\alpha^{2/3}}\frac{\hbar^4c^2p_{F \rm p}}{p_{F \rm e}^2+p_{F\mu}^2} \,
  \left(\frac{\Delta}{p_{F \rm p}c}\right)^{1/3}, \ \ \ \ \
  \xi_S=\frac{27\sqrt{3}\pi^{1/3}}{40}\approx 1.71.
\end{equation}
One can use this expression for estimates, but
we recommend to employ the total collision frequency
in practical calculations.

\subsection{Neutron shear viscosity in superfluid matter}

We study the effect of proton superfluidity on neutron-proton
collisions. Even this problem is difficult and
we adopt a simplified approach used in BHY01
for the problem of neutron thermal conductivity.
It has also been widely used for analyzing superfluid
suppression of various neutrino processes (e.g., \cite{yls99}
and references therein). It consists in
taking an ordinary differential probability of a given
scattering process (neutron-proton scattering, in our case)
and inserting particle energies (\ref{E:DispersSF}) with
energy gaps in corresponding Fermi-Dirac
distribution functions. In our case, this approach
is expected to be sufficiently accurate.
Let us recall that we
consider the protons only as neutron scatterers.
Proton superfluidity suppresses this scattering channel,
and our approach reproduces such a suppression.

In this approximation, the neutron-proton collision frequency
becomes
\begin{equation}
  \nu_{\rm np}^{S}=\nu_{\rm np} R_{\rm np}(y),
\end{equation}
where $R_{\rm np}(y)$ is the superfluid reduction factor. The
latter factor is given by the same expression as the reduction
factor for lepton-proton collisions, $R^\parallel_{\rm p}$, save the
coherence factors,
\begin{eqnarray}
R_{\rm np}(y)&=&
     \frac{3}{\pi^2}\int\limits_0^\infty\int\limits_0^\infty
     \frac{{\rm d}x_2\,{\rm d}x_{2'}}{1+\exp(z_2)}
     \left\{\frac{z_{2'}-z_2}
     {[1+\exp(-z_{2'})][\exp(z_{2'}-z_2)-1]} \right.\nonumber\\
&-&\left.
     \frac{z_{2'}+z_2}
     {[1+\exp(z_{2'})][\exp(-z_{2'}-z_2)-1]}
     \right\}.
\end{eqnarray}
It obeys the asymptotes $R_{\rm np}(0)=1$ and $R_{\rm np}(y)\to
A_{\rm np} \,y\exp(-y)$ at $y\to\infty$, where $A_{\rm
np}=0.8589$.

We have calculated $R_{\rm np}(y)$ in a wide range of $y$ and
fitted the results by the expression
\begin{eqnarray}
    R_{\rm np}(y)&=&\frac{2}{3}
    \left[ 0.513+\sqrt{(0.487)^2+0.018\,y^2} \right]
    \exp \left[ 2.26-\sqrt{(2.26)^2+y^2} \right]
\nonumber\\
&+&
    \frac{1}{3}\left(1+0.00056\,y^4\right)
    \exp \left[ 6.2-\sqrt{(6.2)^2+4\,y^2} \right] ;
\end{eqnarray}
the formal maximum fit error is $\approx 0.25\%$ at $y=11.6$.

Note, that proton superfluidity affects $\eta_\mathrm{n}$
weaker than $\eta_\mathrm{e\mu}$ (because of a relatively
small contribution of neutron-proton collisions to
$\eta_\mathrm{n}$).

\section{Results and discussion}
\label{results}

\subsection{Equations of state}
\label{S:Eos}

Our results can be
used for a wide
range of EOSs
of the npe$\mu$-matter in neutron star cores.
For illustration, we have selected
five model EOSs. The parameters of these EOSs  are given in Table
\ref{tab:eos}, including the maximum
gravitational mass $M_\mathrm{max}$ of stable stars and
the threshold density $\rho_\mu$ of muon appearance.

\begin{table}[ht]
\caption[]{Parameters of the selected EOSs:
The compression modulus $K_0$
of symmetric saturated nuclear matter;
the muon threshold density $\rho_\mu$; and also the central
density $\rho_{\rm max}$, the mass $M_{\rm max}$ and radius $R_{\rm m}$
of maximum-mass models ($\rho_\mu$ and $\rho_{\rm max}$ are given
in units of $10^{14}$ g~cm$^{-3}$)}
\label{tab:eos}
\begin{center}
\begin{tabular}{  c  c  c  c  c  c}
\hline \hline
 EOS & ~~$K_0$~~ & ~$\rho_{\mu14}$~ & ~$\rho_{\rm max14}$~ & ~$M_{\rm max}$~
 & ~~~$R_{\rm m}$~~~ \\
 &MeV &  &  & $M_\odot$ &
 km \\
\hline \hline
 APR~~  & 237 & 2.28  & 27.6 & 1.923 & 10.31 \\
\hline
  & 120  & 2.55  & 38.6 & 1.468 & 9.18 \\
PAL I  & 180  &  2.55 & 31.4 & 1.738 & 9.92 \\
  & 240  & 2.55  & 26.6 & 1.950 & 10.59 \\
\hline
  & 120  & 2.58  & 35.3 & 1.484 & 9.72 \\
PAL II  & 180  &  2.58 & 29.5 & 1.753 & 10.36 \\
  & 240  & 2.58  & 25.3 & 1.966 & 10.97 \\
\hline
  & 120  & 2.46  & 44.4 & 1.416 & 8.45 \\
PAL III  & 180  &  2.46 & 34.5 & 1.713 & 9.60 \\
  & 240  & 2.46  & 28.6 & 1.910 & 10.12 \\
\hline
  & 120  & 2.50  & 42.0 & 1.438 & 8.75 \\
PAL IV  & 180  &  2.50 & 33.2 & 1.713 & 9.60 \\
  & 240  & 2.50  & 27.8 & 1.927 & 10.32 \\
\hline
\end{tabular}
\end{center}
\end{table}

The APR EOS was constructed by Akmal, Pandharipande, and Ravenhall
\cite{apr98} (their model Argonne V18+$\delta v$+UIX$^*$);
it is often used in the literature.
Specifically, we adopt its convenient parametrization proposed by
Heiselberg and Hjorth-Jensen \cite{hh99} and described as APR~I
by Gusakov et al.\ \cite{gus05}.
It is sufficiently stiff, the maximum neutron star mass
is $M_\mathrm{max}\approx 1.92\, M_{\odot}$
(and the maximum-mass star has circumferential radius of $R_{\rm
m}=10.31$ km),  the muons appear at $\rho_\mu\approx2.28 \times
10^{14}$~g~cm$^{-3}$; see Table \ref{tab:eos}.

The PAL EOSs are convenient semi-analytical
phenomenological EOSs proposed by Prakash, Ainsworth, and
Lattimer \cite{pal88}. They differ by the functional form
of the dependence of the symmetry energy $S$ of dense matter on the
baryon number density $n_b$. This dependence is described
\cite{pal88} by a
function $F(u)$, where $u=n_b/n_0$, $n_0=0.16$~fm$^{-3}$ being the
baryon number density of saturated symmetric matter. For the
PAL EOSs I, II, and III, these functions are $F(u)=u$, $2u^2/(u+1)$,
and $\sqrt{u}$, respectively. The PAL IV EOS belongs to the same
family of EOSs, but with the symmetry energy
$S(u) \propto u^{0.7}$ suggested
by Page and Applegate \cite{pa92}. The PAL EOSs differ also  \cite{pal88}
by the value of the compression modulus $K_0$ of saturated
symmetric matter, $K_0=120$, $180$, and
$240$~MeV. Nevertheless, the particle fractions
$n_i/n_b$ as a function of $n_b$ are independent of
$K_0$ (for these EOSs); the dependence
of $n_b$ on $\rho$ is almost identical for the
three selected $K_0$ values [at a fixed $F(u)$].
Hence, the collision frequencies and
the shear viscosity are independent of
$K_0$. However, taking different $K_0$, one
obtains very different neutron star models
(different mass-radius relations
and $M_{\rm max}$; see Table \ref{tab:eos}). For
illustration, we take $K_0=240$~MeV for all PAL models
(unless the contrary is indicated).

Therefore, our selected EOSs correspond
to a large variety of neutron star models.

\subsection{Shear viscosity in non-superfluid matter}
\label{S:resnorm}

\begin{figure}[ht]
\begin{center}
\includegraphics[width=0.5\textwidth]{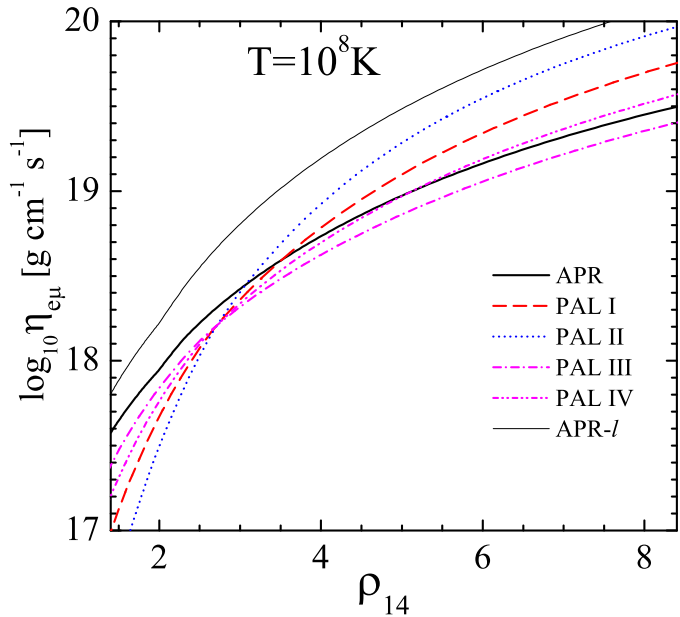}
 \caption{(Color online) Shear viscosity $\eta_\mathrm{e\mu}$
 of electrons and muons versus density $\rho_{14}$
 (in units of $10^{14}$ g~cm$^{-3}$)
 for different EOSs (Table \ref{tab:eos})
 at $T=10^8$~K
 ($m_{\rm p}^*=0.8\,m_{\rm n}$). The thin solid line (APR-$l$) shows the viscosity
 $\eta_\mathrm{e\mu}$ calculated with account for the exchange of
 longitudinal plasmons alone.
} \label{fig:viscem}
\end{center}
\end{figure}

Figure \ref{fig:viscem} shows the shear viscosity
$\eta_\mathrm{e\mu}$ of electrons and muons versus density at $T=10^8$~K
for five EOSs. The given temperature is typical for
middle-aged ($t\sim 10^4-10^5$~yr) isolated (cooling) neutron stars without
enhanced neutrino emission in their cores
(e.g., Refs.\ \cite{pa92,yls99}). The proton effective mass is
taken to be $m_{\rm p}^*=0.8\,m_{\rm n}$.
The thick lines give
$\eta_\mathrm{e\mu}$ for the APR and PAL~I--IV EOSs, while the thin solid
line is for the APR EOS, but it is calculated
including the contribution from the exchange of longitudinal plasmons
alone. One can see, that the inclusion of transverse plasmons
lowers $\eta_\mathrm{e\mu}$ by a factor of three
at $\rho\gtrsim 4\times10^{14}$~g~cm$^{-3}$. With the fall of
temperature this lowering is stronger. The exchange of
transverse plasmons has not been included in previous
calculations of the shear viscosity in neutron star cores, which
has resulted in an overestimation of $\eta_\mathrm{e\mu}$.
The viscosity $\eta_\mathrm{e\mu}$ for the PAL~II EOS (the dotted line)
goes significantly higher than other curves due to a larger amount of
protons (and, therefore, electrons and muons) for this EOS.

\begin{figure}[ht]
\begin{center}
\includegraphics[width=0.55\textwidth]{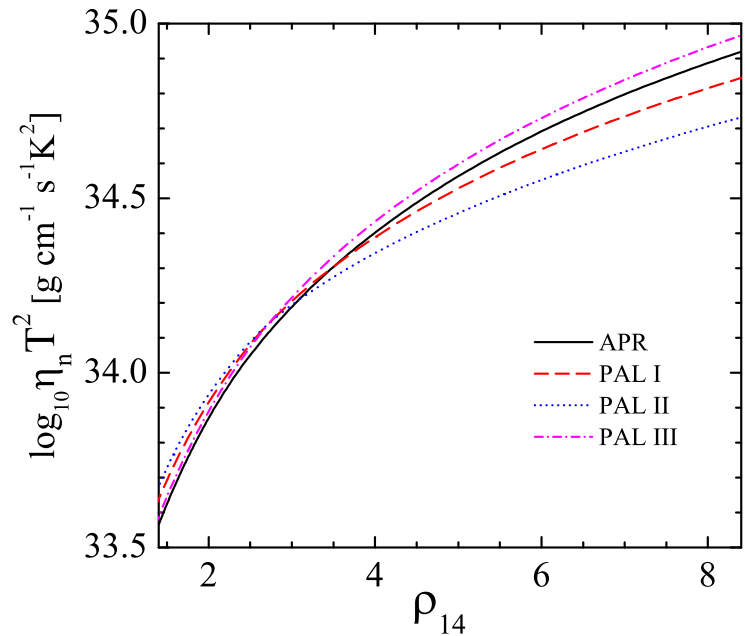}
 \caption{(Color online) Viscosity $\eta_\mathrm{n}$ of neutrons times
 $T^2$ versus density for four EOSs
 ($m^*_\mathrm{n}=m^*_\mathrm{p}=0.8 \, m_\mathrm{n}$).
} \label{fig:viscn}
\end{center}
\end{figure}

Figure \ref{fig:viscn} demonstrates the density dependence of the
neutron shear viscosity multiplied by
squared temperature, $\eta_\mathrm{n}\,T^2$. This combination
is temperature independent. The curves
are calculated assuming
the nucleon effective masses
$m^*_\mathrm{n}=m^*_\mathrm{p}=0.8 \, m_\mathrm{n}$.
In principle, the effective masses can be taken from
microscopic calculations of an EOS; they can depend on
$\rho$, and our expressions for the shear viscosity allow
one to incorporate this density dependence. Here we assume
density independent effective masses by way of illustration.
In Fig.~\ref{fig:viscn}, for simplicity,
we do not present the results for the PAL~IV EOS;
they are very close to the
APR results. One can see, that the neutron viscosity for the
selected EOSs differs within a factor of $\lesssim 2$.

\begin{figure}[ht]
\begin{center}
\includegraphics[width=0.7\textwidth]{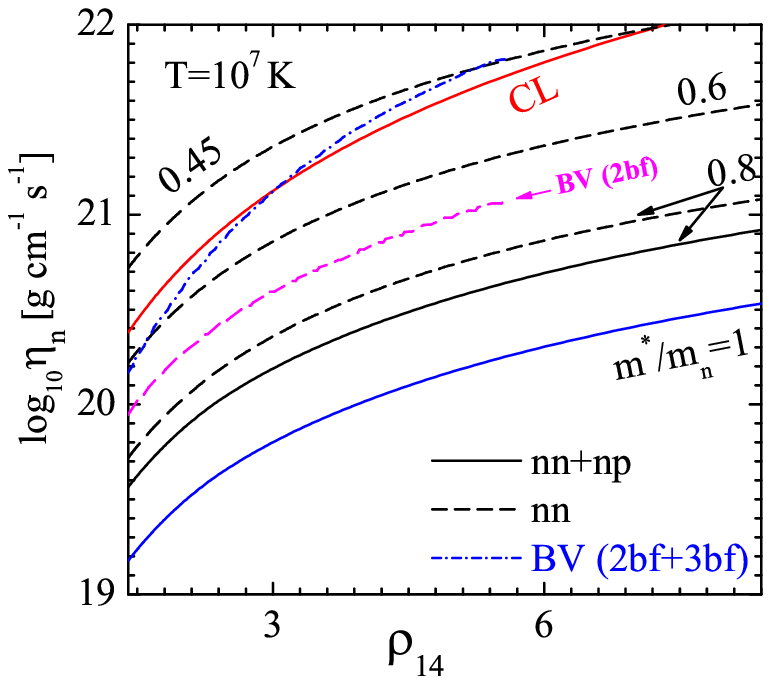}
 \caption{(Color online)
Neutron shear viscosity versus $\rho$ at $T=10^7$~K. The curve CL is
the approximation of Cutler and Lindblom \cite{cl87}
of the results \cite{fi79}. The curves
BV (2bf) and BV (2bf+3bf) are obtained by Benhar and Valli \cite{bv07}
for pure neutron matter
taking into account two-body and two-body plus
three-body forces, respectively.
Other curves are our results for the viscosity $\eta_{\rm nn}$,
limited by neutron-neutron collisions alone, or for the viscosity
$\eta_{\rm n}$, limited by neutron-neutron and neutron-proton
collisions; these curves are calculated for the APR EOS assuming
various phenomenological
density independent effective masses $m_{\rm n}^*$ and $m_{\rm p}^*$.
} \label{fig:BV}
\end{center}
\end{figure}

Figure \ref{fig:BV} demonstrates the
viscosity approximation of Cutler and Lindblom \cite{cl87} (curve CL)
versus $\rho$ at $T=10^7$~K.
Recall that the approximation is based on the calculations
by Flowers and Itoh \cite{fi79} performed for the EOS of Baym,
Bethe and Pethick \cite{bbp71} assuming in-vacuum nucleon-nucleon
scattering cross sections and
$m^*_\mathrm{n}=m^*_\mathrm{p}=m_\mathrm{n}$. Also, we show
self-consistent calculations of $\eta_\mathrm{n}$
by Benhar and Valli \cite{bv07}
for a pure neutron matter with
the EOS that is basically similar to APR (with another version for
three-nucleon interaction). The authors have used one and the same
nucleon interaction potential to derive the EOS and
$\eta_\mathrm{n}$. The curve BV (2bf)
is their result  (from their Fig.~1)
obtained employing two-body nucleon forces;
the curve BV (2bf+3bf) is obtained employing
the two-body and three-body forces.
The effective mass $m^*_\mathrm{n}$ is calculated self-consistently
as a function of $\rho$ ($m^*_\mathrm{n}$ is different for both curves
and, unfortunately, is not
reported in \cite{bv07}).

All other curves in Fig. \ref{fig:BV} are our results for the APR
EOS assuming various values of $m^*_\mathrm{n}$ and
$m^*_\mathrm{p}$. For simplicity, these phenomenological values
are taken density independent.
We show either the viscosity $\eta_\mathrm{nn}$,
limited by neutron-neutron collisions alone (dashed lines), or the
viscosity $\eta_\mathrm{n}$, limited by neutron-neutron and
neutron-proton collisions (solid lines). One can see that the
contribution of neutron-proton collisions is relatively small,
while the dependence of the viscosity on nucleon effective masses
is important. Smaller effective masses strongly increase the
neutron viscosity. In the limit of
$m^*_\mathrm{n}=m^*_\mathrm{p}=m_\mathrm{n}$ we obtain the
viscosity $\eta_\mathrm{n}$ which is a factor of $\approx 40$
smaller than CL. Using the results of BHY01 for the thermal
conductivity of neutrons $\kappa_{\rm n}$, derived in the same
approximations as our results for $\eta_{\rm n}$, we obtain the
values of $\kappa_{\rm n}$ a factor of 2--4 smaller than those
given by Flowers and Itoh \cite{fi79,fi81}. The nature of this
systematic disagreement of our results with the results of Flowers
and Itoh is unclear. We have checked that it cannot be attributed
to using different EOSs.

A comparison of our results with those of Benhar
and Valli \cite{bv07} is complicated because Benhar and Valli
do not present the
values of $m_\mathrm{n}^*$ which they obtained for a neutron
matter. If, however, we take
a reasonable value of $m_{\rm n}^*=0.7m_{\rm n}$, we obtain $\eta_\mathrm{nn}$
(not shown in Fig.\ \ref{fig:BV})
very close to the curve BV (2bf) of Benhar and Valli.
In order to reproduce their BV (2bf+3bf) curve with our equations, we
should employ a density dependent $m_\mathrm{n}^*$. It should
vary from $m_\mathrm{n}^*\approx 0.6\,m_{\rm n}$ at
$\rho \sim 1.5\times 10^{14}$~g~cm$^{-3}$ to $0.45\,m_{\rm n}$ at
$\rho \sim 6\times 10^{14}$~g~cm$^{-3}$. Let us
note, that the inclusion of three-nucleon interactions does
reduce $m_\mathrm{n}^*$, and the reduction increases
with density \cite{zhangetal07}. However, since we do not know
exact values of $m_\mathrm{n}^*$, used in Ref.~\cite{bv07},
we cannot analyze the relative importance
of $m_\mathrm{n}^*$ and in-medium corrections to the squared matrix
element. We assume  (Sec.~\ref{viscn_norm})
that the effect of the effective masses is
more important.
Notice, in addition, that
the in-vacuum differential neutron-neutron scattering cross section
in Ref.\ \cite{bv07}
(the solid line in their Fig.\ 3) seems underestimated.

\begin{figure}[ht]
\begin{center}
\includegraphics[width=0.5\textwidth]{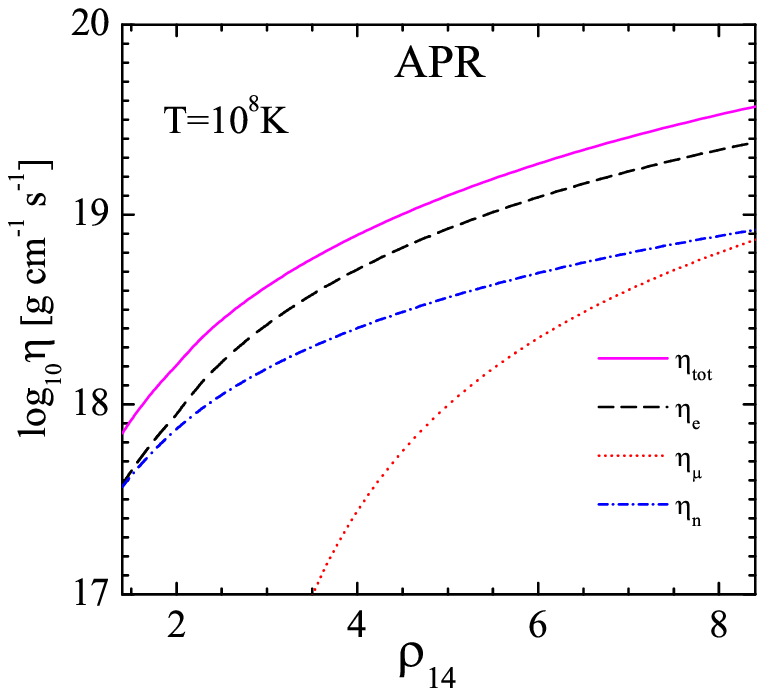}
 \caption{(Color online) Partial shear viscosities in non-superfluid neutron star
 cores versus density at $T=10^8$~K ($m_{\rm n}^*=m_{\rm p}^*=0.8m_{\rm n}$).
} \label{fig:partial}
\end{center}
\end{figure}

In Fig.\ \ref{fig:partial} we compare partial shear viscosities
in a neutron star core with the APR EOS at $T=10^8$~K. Previously, it has
been widely thought that $\eta_\mathrm{n}$ completely dominates
over $\eta_\mathrm{e\mu}$ in the core of a nonsuperfluid
star. Now we have considerably lowered both viscosities (Figs.\
\ref{fig:viscem} and \ref{fig:BV}). The main contribution to the
total shear viscosity ($\eta_{\rm tot}$, the solid line) at
$T=10^8$~K comes from the electrons ($\eta_\mathrm{e}$, the dashed
line). The neutron viscosity $\eta_\mathrm{n}$ (the dash-dotted
line) is lower than $\eta_\mathrm{e}$. Note, however, that the
relation between $\eta_\mathrm{n}$ and $\eta_\mathrm{e}$ is
temperature-dependent; when $T$ decreases, $\eta_\mathrm{n}$
becomes more important (see Fig.\ \ref{fig:viscT} and a
discussion below).
The dotted line
in  Fig.\ \ref{fig:partial} shows the muon shear viscosity
$\eta_\mu$.  For $T=10^8$~K, it becomes comparable with
$\eta_\mathrm{n}$ at $\rho\gtrsim 7\times10^{14}$~g~cm$^{-3}$.

\subsection{Shear viscosity in superfluid matter}
\label{S:ressf}
Now we discuss the shear viscosity in the presence
of proton superfluidity (superconductivity) but for nonsuperfluid neutrons.
For illustration, we take $m^*_\mathrm{n}=m^*_\mathrm{p}=0.8 \, m_\mathrm{n}$
throughout a neutron star core.

\begin{figure}[ht]
\begin{center}
\includegraphics[width=0.5\textwidth]{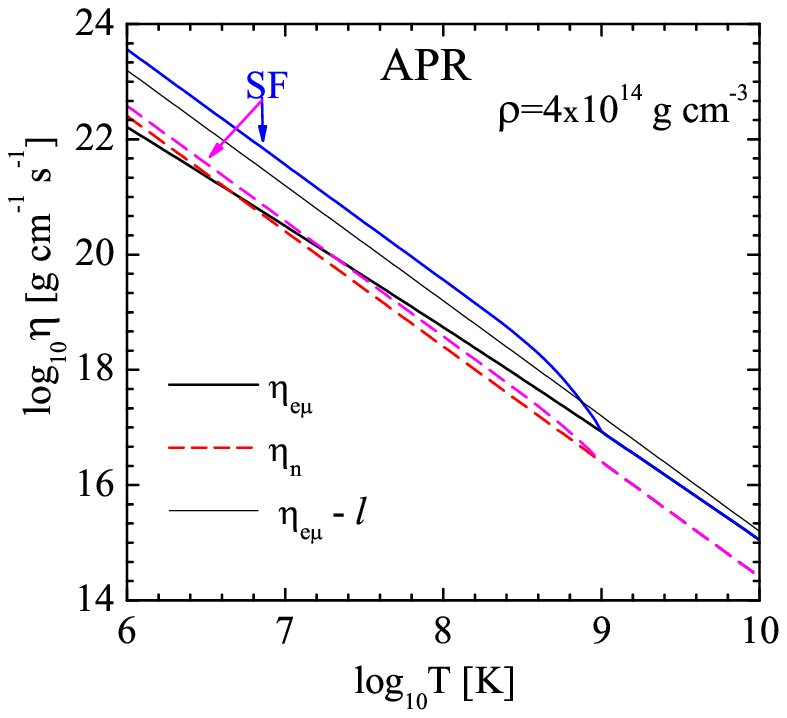}
 \caption{(Color online) Electron-muon and neutron shear viscosities
versus temperature in the
non-superfluid neutron star core and in the presence of proton superfluidity
($T_{c \rm p}=10^9$~K)
for the APR EOS at $\rho=4\times 10^{14}$~g~cm$^{-3}$
($m_{\rm n}^*=m_{\rm p}^*=0.8m_{\rm n}$).
Curves SF correspond to the superfluid case, while other
curves are for normal matter. The thin solid curve is
$\eta_\mathrm{e\mu}$ calculated including the exchange of longitudinal
plasmons alone.} \label{fig:viscT}
\end{center}
\end{figure}

Figure \ref{fig:viscT} demonstrates the temperature dependence of
$\eta_\mathrm{e\mu}$ (solid lines)
and $\eta_\mathrm{n}$ (dashed lines) in the presence of proton superfluid
($T_{c \rm p}=10^9$~K, curves SF) and for
non-superfluid matter (unmarked curves) at
$\rho=4\times10^{14}$~g~cm$^{-3}$.

In a non-superfluid matter, $\eta_\mathrm{e\mu}$  exceeds
$\eta_\mathrm{n}$ at $T\gtrsim 10^7$~K (for the adopted values of
$m_{\rm n}^*$ and $m_{\rm p}^*$) but the situation reverses at
lower $T$. The reversal is a consequence of the different
temperature behaviors, $\eta_\mathrm{e\mu}\propto T^{-5/3}$
[Eq.\ (\ref{eq:etaperp})] and $\eta_\mathrm{n}\propto T^{-2}$. The
thin solid line shows $\eta_\mathrm{e\mu}$ calculated taking into
account the exchange of longitudinal plasmons alone. It
demonstrates the standard Fermi-system behavior,
$\eta_\mathrm{e\mu}\propto T^{-2}$, and overestimates
$\eta_\mathrm{e\mu}$. At $T=10^9$~K
the overestimation 
is small. It reaches a factor of $\sim$ three at $T =
10^8$~K, and exceeds one order of magnitude at $T\lesssim 10^7$~K.

Proton superfluidity noticeably increases
$\eta_\mathrm{e\mu}$ at $T<T_{c \rm p}$ and  restores the Fermi-liquid
temperature behavior, $\eta_\mathrm{e\mu}\propto T^{-2}$
[Eq.\ (\ref{etaper_sf})]. The increase of $\eta_\mathrm{n}$ is not
large because it comes from superfluid suppression of
neutron-proton collisions which give a relatively small contribution
to $\eta_\mathrm{n}$. In the presence of proton superfluidity,
$\eta_\mathrm{e\mu}$ completely dominates over $\eta_\mathrm{n}$.

The electron shear viscosity in superfluid matter has recently
been considered by Andersson et al.\ \cite{acg05}. We have already
discussed their approach in Sec.\ \ref{red_prot}. For their
EOS and superfluidity model, their results
overestimate $\eta_\mathrm{e}$,
typically, by a factor of three in superfluid matter
and by more than one order of magnitude in non-superfluid
matter.

\begin{figure}[th]
\begin{center}
\includegraphics[width=1.0\textwidth]{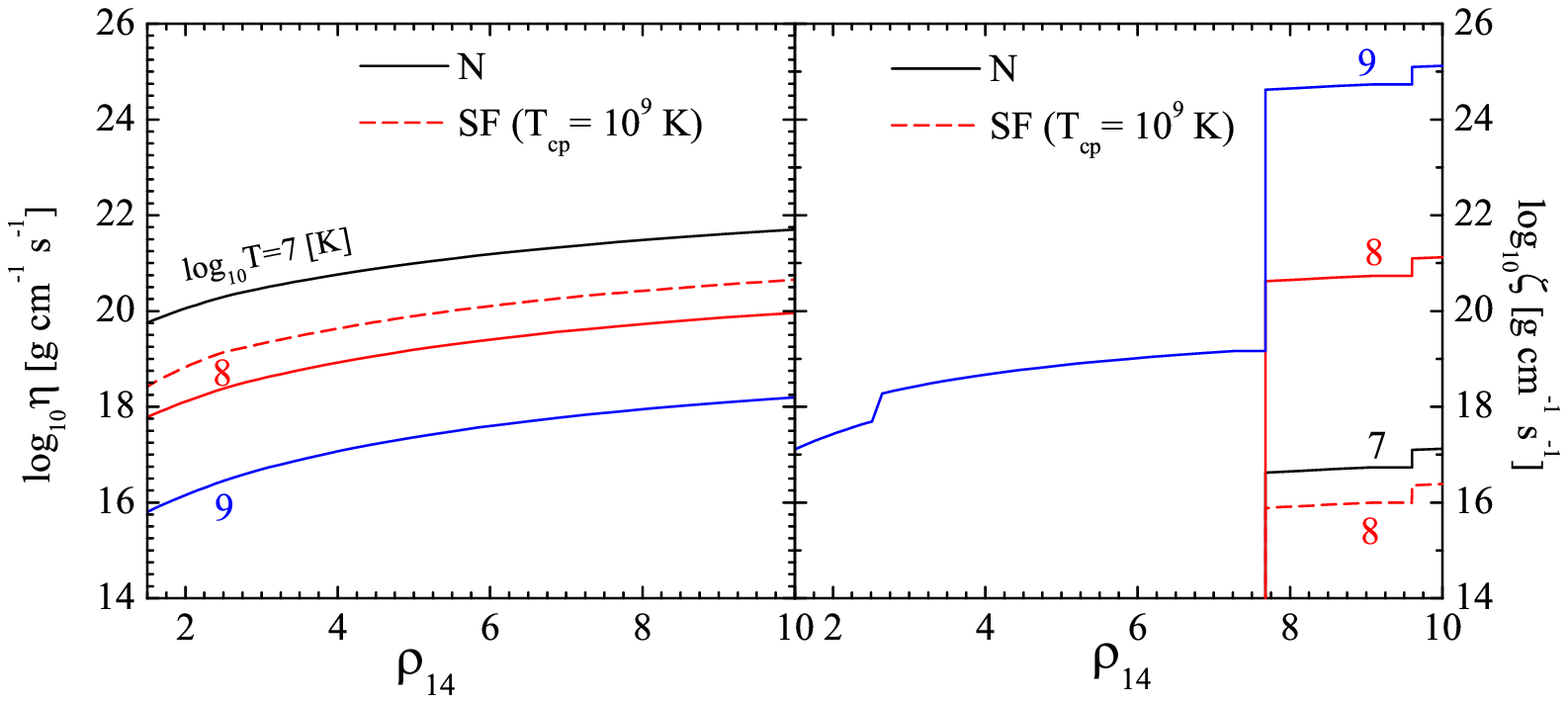}
 \caption{(Color online) Density dependence of the shear viscosity $\eta$
 (left) and the bulk viscosity
 $\zeta$ (right)
 at different temperatures (the values of
 $\log_{10} T$ are given near the curves)
 in a neutron star core with the
 PAL~I EOS (see text).
 Solid curves are for normal matter (N).
 Dashed curves are for
 $T=10^8$~K and proton superfluidity with $T_\mathrm{cp}=10^9$~K.
 The viscosity $\zeta$ is plotted for a neutron star vibrating
 at a frequency of $\omega=10^4$~s$^{-1}$.
} \label{fig:Bulk}
\end{center}
\end{figure}

Figure \ref{fig:Bulk} compares the shear viscosity $\eta$ (left
panel) with the bulk viscosity $\zeta$ (right panel) determined by
the direct and modified Urca processes in the core of a vibrating
neutron star at $T=10^7$, $10^8$, and $10^9$~K. The vibration
frequency is set to be $\omega=10^4$~s$^{-1}$; these vibrations
strongly affect $\zeta$ but do not affect $\eta$. The bulk
viscosity is calculated according to Refs.\ \cite{hly00,hly01}.
The EOS is the same as in \cite{hly00,hly01} (PAL~I with
$K_0=180$~MeV).

Note, that $\eta$ decreases with growing $T$, while $\zeta$
increases (e.g., Refs.\ \cite{hly00,hly01}). For $T=10^7$~K, the
shear viscosity dominates in the entire stellar core, while for
$T=10^8$~K the bulk viscosity $\zeta$ in the inner core (where the
direct Urca process is allowed, after the jump of $\zeta$
in the right panel) becomes $\sim$10 times higher than
$\eta$. For $T=10^9$~K, the bulk viscosity completely dominates in
the entire core. The presence of proton superfluidity enhances
$\eta$ and suppresses $\zeta$. The dashed lines in the left and right
panels of Fig.\ \ref{fig:Bulk} show $\eta$ and $\zeta$,
respectively, in superfluid matter with $T_\mathrm{cp}=10^9$~K at
$T=10^8$~K. Superfluidity makes the shear viscosity more
important. Note that shear perturbations in dense matter (e.g.,
associated with differential stellar rotation) are damped by the
shear viscosity and can be unaffected by the bulk viscosity.
Therefore, the shear viscosity can be important for applications
even if it is lower than the bulk viscosity.

\begin{figure}[ht]
\begin{center}
\includegraphics[width=0.5\textwidth]{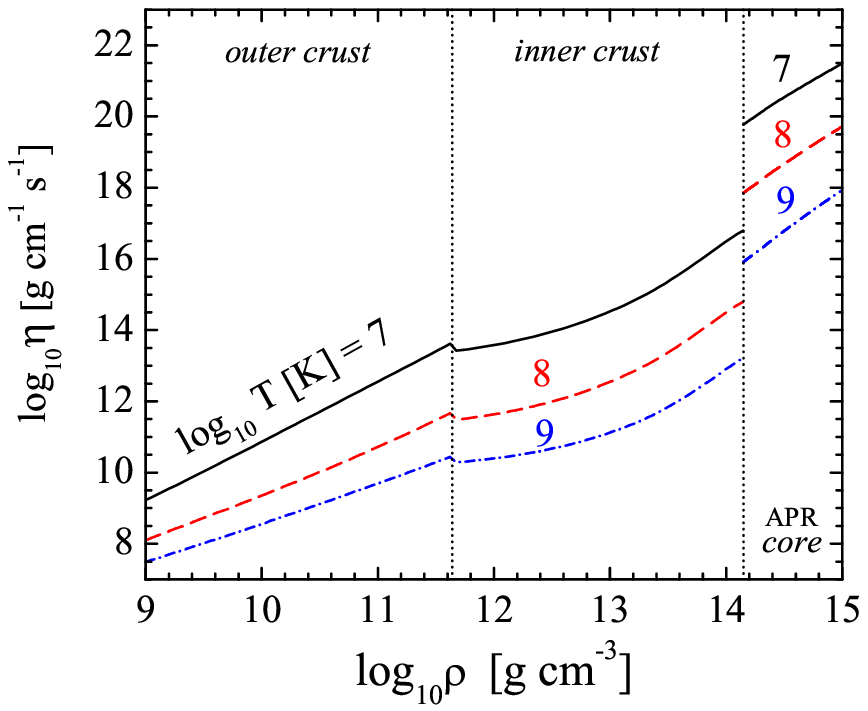}
 \caption{(Color online) Density profiles of the shear viscosity through
 a non-superfluid neutron
 star (through the outer crust, inner crust and the core)
 for three values of $T$
 ($\log_{10} T$~[K] =7, 8, and 9 -- solid, dashed,
 and dash-dotted lines, respectively).
 The left vertical dotted line shows the neutron
 drip density, the right line is the crust-core
 interface.}
 \label{fig:CoreCrust}
\end{center}
\end{figure}

Finally, Fig.\ \ref{fig:CoreCrust}
compares the shear viscosity in the crust and the core of a
neutron star. The viscosity in the crust is
calculated for  cold-catalyzed matter \cite{hpy07}
using the results of Refs.\ \cite{chy05} and \cite{sh08a}. The former paper
is devoted to the viscosity mediated by
electron-ion collisions, while Ref.\
\cite{sh08a} deals with the contribution of electron-electron collisions taking
into account the exchange of transverse plasmons.
In the crust, the latter effect
is not large. 
We use the APR EOS in the core, and the core is assumed to be
non-superfluid. The solid, dashed, and dash-dot lines
correspond to $T=10^7$, $10^8$, and
$10^9$~K, respectively. The viscosity jump at the
star crust-core interface (at
$\rho=1.4\times10^{14}$~g~cm$^{-3}$) is due to the disappearance of
atomic nuclei in the core. The nuclei, present in the crust,
lower the viscosity owing to a very efficient electron-ion scattering.

\section{Conclusions}

We have calculated the shear viscosity in a neutron star core
as a sum of the electron-muon viscosity
$\eta_\mathrm{e\mu}$ and the neutron viscosity
$\eta_\mathrm{n}$. Calculating the viscosity $\eta_\mathrm{e\mu}$,
which is mediated by collisions of charged particles, we have taken into
account the exchange of transverse plasmons
(that has not been done before).
Our results include also the effects of proton superfluidity. They
are universal, presented in the form of analytic fit
expressions convenient for implementing into computer codes for
any EOS of nucleon matter in neutron star
cores.

Our main conclusions are:
\begin{enumerate}
\item The exchange of transverse plasmons strongly reduces
$\eta_\mathrm{e\mu}$ for all temperatures and densities of
interest in a non-superfluid core. A low temperatures, we
have $\eta_\mathrm{e\mu}\propto T^{-5/3}$.

\item The  viscosity $\eta_\mathrm{e\mu}$ generally dominates over
$\eta_\mathrm{n}$, although $\eta_\mathrm{n}$ can exceed
$\eta_\mathrm{e\mu}$ at  $T\lesssim 10^7$~K and $\rho\lesssim
4\times 10^{14}$~g~cm$^{-3}$ (for $m_{\rm n}^*\approx m_{\rm p}^*\approx 0.8m_{\rm n}$).

\item The viscosity  $\eta_{\rm n}$ strongly depends on the
nucleon effective masses. Typically, it is more than one order of
magnitude lower, than that calculated in Ref.\ \cite{fi79} and
parametrized in
Ref.\ \cite{cl87}.

\item Strong proton superfluidity significantly increases
$\eta_\mathrm{e\mu}$ and restores its Fermi-liquid
temperature dependence,
$\eta_\mathrm{e\mu}\propto T^{-2}$. In this regime, $\eta_\mathrm{e\mu}$
exceeds $\eta_\mathrm{n}$.

\item The shear viscosity $\eta$ is comparable with the bulk viscosity
$\zeta$ at $T \sim 10^8$~K (for a star vibrating
at a frequency
$\omega \sim 10^4$~s$^{-1}$) and dominates at lower $T$.
Superfluidity increases the importance of $\eta$ in comparison
with $\zeta$.
\end{enumerate}

Our results can be used in simulations of neutron star
hydrodynamics, in particular, to analyze the damping of
internal differential rotation, stellar oscillations,
gravitational wave driving instabilities.

Our results can be improved further in many respects.
It would be most important to
study the shear viscosity problem in the presence of
neutron and proton superfluidity in the frame of
multifluid hydrodynamics as discussed in Sec.\ \ref{superflu}.

Nevertheless, even our restricted standard one-fluid
formulation is incomplete.
Our calculations of $\eta_{\rm n}$ can be improved by taking into
account the medium effects on the matrix elements of
nucleon-nucleon scattering. However, we rely on the results of
Ref.\ \cite{zhangetal07} that these medium effects are  weaker
than the effects of nucleon effective masses
(which we include explicitly). An account for the medium effects on
the matrix elements would complicate the expressions for $\eta_{\rm n}$
(making them non-universal).

We have also neglected the effects of strong magnetic
field which can modify the shear viscosity. For
not too high magnetic fields, $B \lesssim 10^{13}$~G, which do not affect
the plasma polarization
functions (e.g., Ref.\ \cite{sh08b}), the generalization of the present
results to the magnetic case is straightforward.
For stronger fields, the polarization tensor
becomes anisotropic and the viscosity problem is very complicated.

The present results are in line with our studies of kinetic
properties of relativistic plasma taking into account the
exchange of transverse plasmons. These effects were studied
by Heiselberg and Pethick \cite{hp93} for
ultra-relativistic quark plasma. They should be included in all
calculations of kinetic properties of relativistic
plasmas, particularly in neutron stars.  For the neutron star
crust, the effect was studied in \cite{sy06} (thermal
conductivity) and \cite{sh08a} (shear viscosity). For
neutron star cores, it was analyzed in \cite{sy07} (thermal conductivity),
\cite{sh08b} (electrical conductivity), and \cite{jgp05}
(neutrino pair emission in electron-electron collisions).

\begin{acknowledgments}
We are very grateful to U.~Lombardo,
A.~I.\ Chugunov and D.~A.\ Baiko for useful discussions and
critical remarks. This work was partly supported by the Dynasty
Foundation, by the Russian Foundation for Basic Research (grants
08-02-00837, 05-02-22003), and by the State Program ``Leading
Scientific Schools of Russian Federation'' (grant NSh 2600.2008.2).
\end{acknowledgments}

\begin{appendix}
\section{Explicit expressions for angular integrals}
\label{appendix}

Here we present explicit expressions of the angular
integrals $I^\parallel_k(x)$, defined by Eq.\ (\ref{E:parint}), for
different values of $k$,
\begin{eqnarray}
   I^\parallel_{0}(x)&=&\frac{1}{2}\arctan x +\frac{1}{2}\,
   \frac{x}{1+x^2}, \\
   I^\parallel_{2}(x)&=&\frac{1}{2}\arctan x -\frac{1}{2}\,
   \frac{x}{1+x^2},\\
   I^\parallel_{4}(x)&=&x-\frac{3}{2}\arctan x
   +\frac{1}{2}\,\frac{x}{1+x^2},\\
   I^\parallel_{6}(x)&=& \frac{x^3}{3}-2x + \frac{5}{2}\arctan x
   -\frac{1}{2}\,\frac{x}{1+x^2},\\
   I^\parallel_{8}(x)&=& \frac{x^5}{5}-\frac{2x^3}{3}+3x - \frac{7}{2}\arctan x
     +\frac{1}{2}\,\frac{x}{1+x^2}.
\end{eqnarray}

%

\end{appendix}

\end{document}